\documentclass[aps,prd,showpacs,preprint,preprintnumbers,showpacs,showkeys,superscriptaddress,amsmath,amssymb,nofootinbib]{revtex4-1}
\usepackage{amsmath}
\usepackage{latexsym}
\usepackage{pst-all}
\usepackage{color}
\usepackage{latexsym}
\usepackage{amsmath}
\usepackage{amssymb}
\usepackage{eufrak}
\usepackage{euscript}
\usepackage[latin1]{inputenc}
\usepackage{pstricks}
\usepackage{graphics}
\usepackage{graphicx}
\usepackage{picture}

\newcommand{\be}{\begin{equation}}
\newcommand{\ee}{\end{equation}}
\newcommand{\ba}{\begin{eqnarray}}
\newcommand{\ea}{\end{eqnarray}}

\def\uma{\rm 1\!\!\hskip 1 pt l}
\begin{document}

\markright{The CP violation and scalar dark matter in a 331 model}
\title{The CP violation and scalar dark matter in a 331 model}
\author{M. J. Neves}
\email{mariojr@ufrrj.br}
\affiliation{Department of Physics and Astronomy, University of Alabama, Tuscaloosa, Alabama 35487, USA
\\
and Departamento de F\'isica, Universidade Federal Rural do Rio de Janeiro, Serop\'edica, RJ 23890-971, Brazil}

\date{\today}



\begin{abstract}
\noindent

The 331 model with right-handed neutrinos is re-assessed to investigate
the CP violation in the quark sector. After the spontaneous
symmetry breaking, the masses and physical fields of the particle content are obtained.
The fermions content of the 331 model is enlarged to include exotic quarks with known electric charge
and with masses defined at the TeV scale. The existence of these exotic quarks induces extra CP violations via
couplings with quarks of the Standard Model mediated by charged gauge boson with mass fixed at
the TeV scale. An extra discrete $\mathbb{Z}_{2}$ symmetry is introduced in the 331 model to get a stable scalar field
that can be a candidate to the dark matter content. The new scalar field interacts at tree level with the $Z^{\prime}$
gauge boson that works as a dark matter portal. The relic density associated with the scalar field is calculated to yield the solution mass
that satisfies the observed dark matter. The region allowed on the parameter space of the dark matter mass versus $Z^{\prime}$ mass
is obtained to include the bounds of PANDAX2017, XENON1T(2t.y) and LUX experiments.
%

%
%
\end{abstract}

\pacs{11.15.-q, 11.10.Nx, 12.60.-i}
\keywords{Physics beyond the Standard Model, CP violation, scalar dark matter.
}

\maketitle

\pagestyle{myheadings}
\markright{The CP violation and scalar dark matter in a 331 model}
%
%
%
%
%
%
%

\section{Introduction}
%

%
%

%
The Standard Model (SM) is the most complete framework to explain the unification of three fundamental interactions,
{\it i.e.}, the unification of the Electroweak (EW) and the Strong interactions.
However, experimental results point out that the SM must be part of more fundamental and complete theory when the particle physics
is investigated at the TeV scale range \cite{CMS,ATLAS}. The measurements of neutrino oscillations show the transition probabilities are associated
with the difference of their squared masses \cite{BellPRL2005}. Thereby, the SM must be extended to contain massive neutrinos.
The simplest extension of the SM introduces the gauged extra group $U(1)_{B-L}$ (baryon minus lepton number)
as a viable inclusion of right-handed neutrinos (RHNs) that acquire masses via type I, II and III seesaw mechanism \cite{seesaw1,seesaw2,seesaw3,seesaw4,seesaw5,VallePRD1980,RabiPRD81,Lazarides81,Wetterich81,Buchmuller91,Rabi98}.
For a general review of the SM extension with a extra gauged group $U(1)$, see \cite{LangackerRMP2009}.

Another extension of the SM known in the literature is the $331$ model. This model is gauged by
$SU_{c}(3) \times SU_{L}(3) \times U(1)_{X}$, in which the $SU_{L}(2)$ group of the EW model is substituted
by $SU_{L}(3)$, and the abelian sector is set by $U(1)_{X}$. The early motivation for the 331 model is to study
the unification schemes of gauge groups for the EW interaction \cite{Ueda73}. The consequence is the enlargement of
the fermions sector to include multiples with exotic quarks and left-handed Majorana neutrinos, the gauge sector is extended
to contain new charged and neutral gauge bosons, and the Higgs sector includes new scalar multiplets \cite{Singer74,SingerPRD80}.
As example of the phenomenology, the 331 model explains processes like $e^{-}e^{-} \rightarrow W^{-} V^{-}$ induced by right-handed
currents coupled to the new charged gauge boson $V^{-}$ \cite{PisanoPRD92}. Nowadays, the 331 model in many different versions is applied
to the particle physics phenomenology at the LHC. A huge number of papers has been explored as a $Z^{\prime}$ benchmark model in the literature \cite{Ninh2005,Cao2015,SanchezVega,Rojas2019,Cao2016,Huitu,Doff,Singh,corcella2019,barreto2019,nomura2019,Dias2005,MonteroPRD2018,EstebanJHEP2019,MonteroPRD15}.
%
%

%
%
The explanation for the dark matter (DM) content in the universe today is another motivation for the particle physics beyond the SM.
A huge number of models has been studied in the literature to include all types of candidates to the DM content like scalars, fermions, and gauge bosons \cite{AmitEPJC2017,PerezPRD2018,FarinaldoReview,Yaguna2015,NandaPRD2017,MandalPRD,Nobu2010}. The SM extension with the extra $U(1)_{X}$ is one of the
most well motivated models to include Dirac fermions candidate as a DM content \cite{FarinaldoJHEP2015,FarinaldoPRD2015,HanWanEPJC2018}.
The $U(1)_{B-L}$ added to a $\mathbb{Z}_{2}$ discrete symmetry includes stable Majorana RHNs that can be candidate to a DM content \cite{SatomiOkada,Kanemura2011,OkadaPRD2017,OkadaMPLA,anirbanJHEP2019}. In the 331 model, a $\mathbb{Z}_{2}$ discrete symmetry is added
to make one of the scalar physical fields stable and candidate to a DM content \cite{Cogollo}. In all these cases, a neutral gauge boson $Z^{\prime}$ or scalar fields with masses at the range $10$ GeV - $10$ TeV works as a DM portal connecting the SM fermions with the DM content.
In this paper, the 331 model $SU_{c}(3) \times SU_{L}(3) \times U(1)_{X}$ is studied with RHNs and added to a discrete symmetry $\mathbb{Z}_{2}$.
The model includes an enlarge fermion sector with three new exotic quarks and one conjugate left-handed neutrino in the multiplets of leptons.
The Higgs sector is extended to contain three scalar multiplets whose the vacuum expected values (VEVs) scales yield masses for all particle content via spontaneous symmetry breaking (SSB). The mechanism is similar to the two-Higgs doublets models (2HDs). The model includes new heavy gauge bosons in which the LHC constraints their masses at the TeV scale. In the basis of the physical fields, the couplings of the exotic quarks with the charged gauge boson are investigated whose phases imply into extra CP violation. The couplings of the physical scalar fields with $Z^{\prime}$ are obtained in the Higgs sector. The discrete symmetry ${\mathbb Z}_{2}$ guarantees the stability of physical scalar fields in which it is the candidate to the DM content of the 331 model. The DM relic density is calculated as function of the DM scalar mass at the TeV scale for the process DM-anti-DM scalar annihilated into SM fermion-anti-fermion pairs through the $Z'$ portal. The parameter space of DM mass versus the $Z^{\prime}$ mass is obtained reproducing
the observed relic density and including the bounds of PANDAX2017 \cite{PANDAX2017}, XENON1T(2t.y) \cite{Xenon20172ty} and LUX \cite{LUXLZ2017} experiments.
The paper is organized as follow the outline : In section \ref{sec2}, the general content of 331 model is presented with RHNs
and the discrete symmetry ${\mathbb Z}_{2}$. The section \ref{sec3} is dedicated to obtain the mass eigenstates for the gauge bosons
and couplings with the fermion content of the 331 model. In section \ref{sec4}, the sector of mass for fermion content and the CP violation
are discussed. The section \ref{sec5} is dedicated to obtain the physical scalar fields in the Higgs sector and their couplings with the $Z^{\prime}$ gauge boson. In section \ref{sec6}, the relic density for the scalar DM candidate and the parameter space with experimental bounds are showed. Finally, our concluding comments are cast in section \ref{sec7}.

%





\section{The 331 model content and the Higgs sector}
\label{sec2}
The model structure is based on the gauge group ${\cal G}_{\rm 331} \equiv SU_{c}(3) \times SU_{L}(3) \times U(1)_X$.
The leptons and quarks are assigned to the following irreducible representations under ${\cal G}_{\rm 331}$ :
\begin{eqnarray}\label{lrSM}
\psi_{iL}  &=&   \left(\begin{array}{c}\nu_{i} \\ e_{i} \\ \nu_{i}^{\, c} \end{array}\right)_{L} \, : \, \left({\bf 1}, {\bf 3}, -\frac{1}{3} \right) \; , \;
Q_{L} = \left(\begin{array}{c} u_{1} \\ d_{1} \\ u_{4}  \end{array}\right)_{L} \, : \, \left({ \bf 3}, {\bf 3}, +\frac{1}{3} \right)
\; , \;
\nonumber \\
q_{aL} &=& \left(\begin{array}{c} d_{a} \\ u_{a} \\ d_{a+2}  \end{array}\right)_{L} \, : \, \left({ \bf 3}, {\bf 3}^{\ast}, 0 \right)
\; , \;
e_{iR} \,  :  \, \left({ \bf 1}, {\bf 1}, -1\right) \; , \;
u_{sR} \, : \, \left({ \bf 3}, {\bf 1},+\frac{2}{3}\right) \; , \;
\nonumber \\
d_{tR} \, &:& \, \left({ \bf 3}, {\bf 1},-\frac{1}{3}\right) \; , \;
N_{iR} \, : \, \left({ \bf 1}, {\bf 1}, 0 \right) \; , \;
\end{eqnarray}
where $i= \left\{ \, 1 \, , \, 2 \, , \, 3 \, \right\}$ is the generation index of leptons family, $a=\left\{ \, 2 \, , \, 3 \, \right\}$
are the index for second and third quarks generation, and the right-handed quarks $u_{sR}$ and $d_{tR}$ have index running on
$s=\left\{ \, 1 \, , \, \ldots \, , \, 4 \, \right\}$ and $t=\left\{ \, 1 \, , \, \ldots \, , \, 5 \, \right\}$. The $u_{4}$,
$d_{4}$ and $d_{5}$ are exotic quarks present on the 331 model that provide new phenomenology for particle physics. The left-handed neutrinos
and leptons are accommodate in $SU_{L}(3)$ triplet with the conjugate left-handed neutrino field $\nu_{L}^{c}$. The left-handed quarks $u_{1L}$
and $d_{1L}$ of the first generation are in another $SU_{L}(3)$ triplet together with the exotic quark $u_{4L}$. The $SU_{L}(3)$ triplet $q_{aL}$
contains the quarks of the second and third generation with the LH exotic quarks $d_{4L}$ and $d_{5L}$. By construction, all the right-handed
components (leptons, quarks and neutrinos) are singlets of $SU_{L}(3)$.

%
The fermion sector is described by the Lagrangian :
\begin{equation}\label{Lleptons}
{\cal L}_{f}=\overline{\psi}_{iL} \, i \, \slash{\!\!\!\!D} \, \psi_{iL}
+\overline{e}_{iR} \, i \, \slash{\!\!\!\!D} \, e_{iR}
+\overline{Q}_{L} \, i \, \slash{\!\!\!\!D} \, Q_{L}
+\overline{q}_{aL} \, i \, \slash{\!\!\!\!D} \, q_{aL}
+\overline{u}_{sR} \, i \, \slash{\!\!\!\!D} \, u_{sR}
+\overline{d}_{tR} \, i \, \slash{\!\!\!\!D} \, d_{tR}
\; ,
\end{equation}
in which the covariant derivative operator of ${\cal G}_{331}$ is
\begin{eqnarray}\label{DmuPsi}
D_{\mu} = \partial_{\mu} + i \, g_{c} \, A_{a\mu} \, \frac{\lambda_{a}}{2}
+i \, g_{L} \, W_{aL\mu} \, \frac{\lambda_{aL}}{2} + i \, g_{X} \, B_{\mu} \, X \, \uma \; .
\end{eqnarray}
%
%
The $A_{a\mu}$, $W_{aL\mu} \; \left(a=1,2,...,8\right)$ and $B_{\mu}$ are the gauge fields of $SU_{c}(3)$, $SU_{L}(3)$ and $U(1)_{X}$, respectively.
The $\left\{ \, \lambda_{a}/2 \, , \, \lambda_{aL}/2 \, \right\}$ are the eight Gell-Man matrices, $X$ stands for the generators of $U(1)_{X}$, and
the $g_{c}$, $g_{L}$ and $g_{X}$ are dimensionless coupling constants.
In 331 models, the electric charge content satisfies the relation
\begin{eqnarray}\label{Qem}
Q_{em}=I_{3L}+\beta\,I_{8L}+X \; ,
\end{eqnarray}
where $\beta$ is a real parameter, $I_{3L}:=\lambda_{3L}/2$ and $I_{8L}:=\lambda_{8L}/2$ are the generators of $SU_{L}(3)$ in terms of Gellman's matrices,
$\lambda_{3L}=\mbox{diag}(1,-1,0)$ and $\lambda_{8L}=\mbox{diag}\left(1/\sqrt{3},1/\sqrt{3},-2/\sqrt{3}\right)$.
The solutions for the $\beta$-parameter are $\beta=\pm 1/\sqrt{3}$ such that we choose $\beta=-1/\sqrt{3}$ in this paper.
The particle content of model is resumed in the table (\ref{table1}). The model is anomaly free with this choice of charges.
\begin{table}
\centering
\begin{tabular}{l l l l l}
\hline
\hline
\mbox{Field content} \quad \quad & \quad \quad $I_{3L}$ \quad \quad & \quad \quad $I_{8L}$ \quad \quad & \quad \quad $U(1)_X$ \\
\hline
\hline
$\nu_{iL}$ \quad \quad & \quad \quad $+1/2$ \quad \quad & \quad \quad $+\sqrt{3}/3$ \quad \quad & \quad \quad $-1/3$  \\
$e_{iL}$ \quad \quad & \quad \quad $-1/2$ \quad \quad & \quad \quad $+\sqrt{3}/3$ \quad \quad & \quad \quad $-1/3$  \\
$\nu_{iL}^{c}$ \quad \quad & \quad \quad $0$ \quad \quad & \quad \quad $-\sqrt{3}/3$ \quad \quad & \quad \quad $-1/3$  \\
$e_{iR}$ \quad \quad & \quad \quad $0$ \quad \quad & \quad \quad $0$ \quad \quad & \quad \quad $-1$ \\
$N_{iR}$ \quad \quad & \quad \quad $0$ \quad \quad & \quad \quad $0$ \quad \quad & \quad \quad $0$ \\
\hline
$u_{1L}$ \quad \quad & \quad \quad $+1/2$ \quad \quad & \quad \quad $+\sqrt{3}/6$ \quad \quad & \quad \quad $+1/3$ \\
$d_{1L}$ \quad \quad & \quad \quad $-1/2$ \quad \quad & \quad \quad $+\sqrt{3}/6$ \quad \quad & \quad \quad $+1/3$ \\
$u_{4L}$ \quad \quad & \quad \quad $0$ \quad \quad & \quad \quad $-\sqrt{3}/3$ \quad \quad & \quad \quad $+1/3$ \\
$u_{aL}$ \quad \quad & \quad \quad $+1/2$ \quad \quad & \quad \quad $-\sqrt{3}/6$ \quad \quad & \quad \quad $0$ \\
$d_{aL}$ \quad \quad & \quad \quad $-1/2$ \quad \quad & \quad \quad $-\sqrt{3}/6$ \quad \quad & \quad \quad $0$ \\
$d_{(a+2)L}$ \quad \quad & \quad \quad $0$ \quad \quad & \quad \quad $+\sqrt{3}/3$ \quad \quad & \quad \quad $0$ \\
$u_{sR}$ \quad \quad & \quad \quad $0$ \quad \quad & \quad \quad $0$ \quad \quad & \quad \quad $+2/3$ \\
$d_{tR}$ \quad \quad & \quad \quad $0$ \quad \quad & \quad \quad $0$ \quad \quad & \quad \quad $-1/3$ \\
\hline
$\Phi$ \quad \quad & \quad \quad $+1/2$ \quad \quad & \quad \quad $+\sqrt{3}/6$ \quad \quad & \quad \quad $-1/3$  \\
$\Lambda$ \quad \quad & \quad \quad $-1/2$ \quad \quad & \quad \quad $+\sqrt{3}/6$ \quad \quad & \quad \quad $+2/3$   \\
$\Xi$ \quad \quad & \quad \quad $0$ \quad \quad & \quad \quad $-\sqrt{3}/3$ \quad \quad & \quad \quad $-1/3$   \\
\hline
\hline
\end{tabular}
\caption{ The particle content of 331 model such that the quiral anomaly is cancel out.} \label{table1}
\end{table}
%

%
%
%
%
%
%
%
%
The sector of gauge bosons is described by the Lagrangian :
\begin{equation}\label{Lgauge}
{\cal L}_{gauge}=-\frac{1}{2} \, \mbox{tr}\left(F_{\mu\nu}^{\; 2}\right)
-\frac{1}{2} \, \mbox{tr}\left(W_{L\mu\nu}^{\; 2}\right)
-\frac{1}{4} \, B_{\mu\nu}^{\; 2}
\; ,
\end{equation}
where
the usual field strength tensors components are given by
%
$F_{a\mu\nu}= \partial_{\mu}A_{a\nu}-\partial_{\nu}A_{a\mu}-g_{c} \, f_{abc} \, A_{b\mu} A_{c\nu}$
, $W_{aL\mu\nu}= \partial_{\mu}W_{aL\nu}-\partial_{\nu}W_{aL\mu}-g_{L} \, f_{abc} \, W_{bL\mu} W_{cL\nu}$,
%
$B_{\mu\nu}= \partial_{\mu}B_{\nu}-\partial_{\nu}B_{\mu}$, and the components of usual structure constants $f_{abc}$ of both the groups
$SU_{c}(3)$ and $SU_{L}(3)$:
\begin{eqnarray}
f_{123}=1
\; , \;
f_{147}=-f_{156}=f_{246}=f_{257}=f_{345}=-f_{367}=\frac{1}{2}
\; , \;
f_{458}=f_{678}=\frac{\sqrt{3}}{2} \; .
\end{eqnarray}
The most general Higgs sector invariant under 331 symmetry is governed by the Lagrangian :
\begin{eqnarray}\label{LHiggs}
{\cal L}_{Higgs}&=&
\left(D_{\mu}\Phi\right)^{\dagger} D^{\mu} \Phi
+\left(D_{\mu}\Lambda\right)^{\dagger} D^{\mu} \Lambda
+\left(D_{\mu}\Xi\right)^{\dagger} D^{\mu} \Xi
\nonumber \\
&&
\hspace{-0.4cm}
-\mu_{1}^{2} \, \left(\Phi^{\dagger} \Phi\right) -\lambda_{1} \left(\Phi^{\dagger} \Phi\right)^{2}
-\mu_{2}^{2} \, \left(\Xi^{\dagger} \Xi\right) -\lambda_{2} \left(\Xi^{\dagger} \Xi\right)^{2}
\nonumber \\
&&
\hspace{-0.4cm}
-\mu_{3}^{2} \, \left( \Lambda^{\dagger} \Lambda \right) -\lambda_{3} \left(\Lambda^{\dagger} \Lambda\right)^{2}
-\lambda_{4}\left(\Phi^{\dagger} \Phi\right)\left(\Xi^{\dagger} \Xi\right)
\nonumber \\
&&
\hspace{-0.4cm}
-\lambda_{5}\left(\Phi^{\dagger} \Phi\right)\left(\Lambda^{\dagger} \Lambda\right)
-\lambda_{6}\left(\Xi^{\dagger} \Xi\right)\left(\Lambda^{\dagger} \Lambda\right)
- \lambda_{7} \left(\Phi^{\dagger} \Xi\right)^2
\nonumber \\
&&
\hspace{-0.4cm}
-\lambda_{8}\left(\Phi^{\dagger} \Xi\right)\left(\Xi^{\dagger} \Phi\right)
- \lambda_{9} \left(\Lambda^{\dagger} \Lambda\right)\left(\Phi^{\dagger} \Xi\right)
-\frac{\lambda_{10}}{\sqrt{2}} \, \varepsilon_{ijk} \, \Phi_{i} \, \Lambda_{j} \, \Xi_{k}
\nonumber \\
&&
\hspace{-0.4cm}
- \, f_{ij}^{(e)} \, \overline{\psi}_{iL} \, \Lambda \, e_{jR}
- \, f_{as}^{(u)} \, \overline{q}_{aL} \, \Lambda^{\ast} \, u_{sR}
- \, f_{1t}^{(d)} \, \overline{Q}_{L} \, \Lambda \, d_{tR}
\nonumber \\
&&
\hspace{-0.4cm}
-f_{ij}^{\prime(e)} \, \varepsilon_{i'j'k'} \, \left(\overline{\psi}_{iL}\right)_{i'} \, \left(\psi_{jL}\right)_{j'}^{c}
\, \left(\Lambda^{\ast}\right)_{k'}
- \, f_{ij}^{(N)} \, \overline{\psi}_{iL} \, \Phi \, N_{jR}
\nonumber \\
&&
\hspace{-0.4cm}
- \, f_{at}^{(d)} \, \overline{q}_{aL} \, \Phi^{\ast} \, d_{tR}
- \, f_{1s}^{(u)} \, \overline{Q}_{L} \, \Phi \, u_{sR}
- \, f_{(a+2)t}^{(d)} \, \overline{q}_{aL} \, \Xi^{\ast} \, d_{tR}
\nonumber \\
&&
\hspace{-0.4cm}
- \, f_{4s}^{(u)} \, \overline{Q}_{L} \, \Xi \, u_{sR}
- \, M_{Rij} \, \overline{N^{c}}_{iR} \, N_{jR}
+\mbox{h. c.} \; ,
\end{eqnarray}
where $\left\{ \, \mu_{1} \, , \, \mu_{2} \, , \, \mu_{3} \, \right\}$ and $\lambda_{i} \, \left( \, i=1 \, , \, 2 \, , \, \cdots \, , \, 9 \, \right)$
are dimensionless real parameters, the coupling constant $\lambda_{10}$ has dimension of energy, $f_{ij}^{(e)}$, $f_{ij}^{(N)}$, $f_{as}^{(u)}$, $f_{1t}^{(d)}$, $f_{at}^{(d)}$, $f_{4s}^{(u)}$, $f_{(a+2)t}^{(d)}$, $f_{1s}^{(u)}$ and $f_{ij}^{\prime \, (e)}$ are Yukawa complex coupling constants that yield masses for leptons and quarks of the model. Still in the term of $\lambda_{10}$, $\Lambda_{i}$, $\Phi_{j}$ and $\Xi_{k}$ mean the components $i,j,k=\left\{ \, 1 \, , \, 2 \, , \, 3 \, \right\}$ of each scalar field that will be defined hereafter. It was also introduced a Majorana mass for the RHNs represented by the matrix elements $M_{Rij}$. After the SSB, the neutrinos LHNs and RHNs will acquire mass via seesaw mechanism of type II.
%
%
%
%
%
The scalar multiplets $\Phi$, $\Lambda$, $\Xi$ have the representation
\begin{eqnarray}\label{PhiGaugeparametrization}
\Phi =
\left(
\begin{array}{c}
\phi_{1}^{0} \\
\phi^{-} \\
\phi_{2}^{0} \\
\end{array}
\right) : \left({\bf 1}, {\bf 3} , -\frac{1}{3}\right)
\; , \;
\Lambda &=&
\left(
\begin{array}{c}
\lambda_{1}^{+} \\
\lambda^{0} \\
\lambda_{2}^{+} \\
\end{array}
\right) : \left({\bf 1}, {\bf 3} , +\frac{2}{3}\right)
\; , \;
\Xi=
\left(
\begin{array}{c}
\xi_{1}^{0} \\
\xi^{-} \\
\xi_{2}^{0} \\
\end{array}
\right) : \left({\bf 1}, {\bf 3} , -\frac{1}{3}\right)
\; . \;
%
\end{eqnarray}
%
%
%
The vacuum expected values (VEVs) scales in the Higgs sector are defined by the matrices
\begin{equation}
\langle \Phi \rangle_{0}=
\left(
\begin{array}{c}
\frac{v_{1}}{\sqrt{2}}  \\
0 \\
0 \\
\end{array}
\right)
\hspace{0.3cm} , \hspace{0.3cm}
\langle \Lambda \rangle_{0}=
\left(
\begin{array}{c}
0 \\
\frac{v_{2}}{\sqrt{2}} \\
0 \\
\end{array}
\right)
\hspace{0.3cm} , \hspace{0.3cm}
\langle \Xi \rangle_{0}=
\left(
\begin{array}{c}
0 \\
0 \\
\frac{v_{3}}{\sqrt{2}} \\
\end{array}
\right)
\; , \;
%
\end{equation}
in which it satisfy the SSB pattern condition $v_{3} \gg v=\sqrt{v_{1}^2+v_{2}^2}$, and $v=246 \, \mbox{GeV}$ is the SM VEV scale.
The VEV $v_{3}$ is responsible by symmetry breaking at the TeV scale, and posteriorly, $v_{1}$ and $v_{2}$ break the resting
symmetry at the SM scale of $246$ GeV. The scheme of SSB here is similar to the case of models with two Higgs doublets (2HDs).
The introduction of three VEVs scales is, in minimal, required to insure the masses for all quarks and an appropriate phenomenology \cite{MonteroPRD15}.
%

%
By definition, all the scalar multiplets are singlets of $SU_{c}(3)$, and note that two of three scalar multiplets have the same quantum numbers. Using the Gell-man matrices representation, the minimal coupling of the gauge fields of $SU_{L}(3) \times U(1)_{X}$ with the scalar fields
has the matrix form from the covariant derivative (\ref{DmuPsi})
\begin{equation}\label{matrixMin}
\left(
\begin{array}{ccc}
\frac{g_{L}}{2} \, W_{3L\mu}+\frac{g_{L}}{2\sqrt{3}} W_{8L\mu}+g_{X} X B_{\mu} & \frac{g_L}{\sqrt{2}} \, W_{\mu}^{+} &
\frac{g_{L}}{\sqrt{2}} \, V_{0\mu}
\\
\\
\frac{g_L}{\sqrt{2}} \, W_{\mu}^{-} & - \frac{g_{L}}{2} \, W_{3L\mu}+\frac{g_{L}}{2\sqrt{3}} W_{8L\mu}+g_{X} X B_{\mu} & \frac{g_{L}}{\sqrt{2}} \, Y_{\mu}^{-}
\\
\\
\frac{g_L}{\sqrt{2}} \, \bar{V}_{0\mu} & \frac{g_L}{\sqrt{2}} \, Y_{\mu}^{+} & -\frac{g_{L}}{\sqrt{3}} \, W_{8L\mu}+g_{X} X B_{\mu} \\
\end{array}
\right) \; ,
\end{equation}
where we have defined the $W^{\pm}$ from the SM as
$\sqrt{2} \, W_{\mu}^{\, \, \, \pm}=W_{1L\mu} \, \mp \, i \, W_{2L\mu}$, the new charged gauge bosons at the TeV scale
$\sqrt{2} \, Y_{\mu}^{\, \, \, \pm}=W_{6L\mu} \, \pm \, i \, W_{7L\mu}$, the neutral gauge bosons
$\sqrt{2} \, V_{0}= W_{4L\mu} - i \, W_{5L\mu}$, $\sqrt{2} \, \bar{V}_{0}= W_{4L\mu} + i \, W_{5L\mu}$.
The others neutral gauge bosons $\left\{ \, W_{3L\mu} \, , \, W_{8L\mu} \, , \, B_{\mu} \, \right\}$ mix to yield
the physical gauge bosons $Z$, $Z^{\prime}$ and the photon. Here, $X$ changes depending on the scalar field : $X=-1/3$
if the matrix (\ref{matrixMin}) acts on $\Phi$ and $\Xi$, and $X=+2/3$ if it acts on $\Lambda$.
%
%
%
%

%

%
A discrete symmetry $\mathbb{Z}_{2}$ can be introduced to simplify the 331 model. This means that the exotic quarks
$u_{4R}$, $d_{4R}$, $d_{5R}$, and the scalar triplet $\Xi$ have transformations under $\mathbb{Z}_{2}$ :
\begin{eqnarray}
u_{4R} \, \rightarrow \, -u_{4R}
\hspace{0.5cm} , \hspace{0.5cm}
d_{(4,5)R} \, \rightarrow \, -d_{(4,5)R}
\hspace{0.5cm} , \hspace{0.5cm}
\Xi \, \rightarrow \, - \Xi \; ,
\end{eqnarray}
and the 331 symmetry of the model is modified to $G_{331} \times \mathbb{Z}_{2}$. if we impose this symmetry in the Higgs Lagrangian (\ref{LHiggs}),
the term with $\lambda_{9}$ in the Higgs potential goes to zero, namely, $\left. \lambda_{9}\right|_{\mathbb{Z}_{2}}=0$. Thereby, the Higgs potential with the $\mathbb{Z}_{2}$ symmetry is simplified to yield the expression
\begin{eqnarray}\label{HiggspotentialZ2}
-V_{Higgs}^{\mathbb{Z}_{2}}(\Phi,\Lambda,\Xi)&=&
\mu_{1}^{2} \, \left(\Phi^{\dagger} \Phi\right) +\lambda_{1} \left(\Phi^{\dagger} \Phi\right)^{2}
+\mu_{2}^{2} \, \left( \Lambda^{\dagger} \Lambda \right) +\lambda_{2} \left(\Lambda^{\dagger} \Lambda\right)^{2}
\nonumber \\
&&
\hspace{-0.4cm}
+\mu_{3}^{2} \, \left(\Xi^{\dagger} \Xi\right) +\lambda_{3} \left(\Xi^{\dagger} \Xi\right)^{2}
+\lambda_{4}\left(\Phi^{\dagger} \Phi\right)\left(\Lambda^{\dagger} \Lambda\right)
\nonumber \\
&&
\hspace{-0.4cm}
+\lambda_{5}\left(\Phi^{\dagger} \Phi\right)\left(\Xi^{\dagger} \Xi\right)
+\lambda_{6}\left(\Lambda^{\dagger} \Lambda\right)\left(\Xi^{\dagger} \Xi\right)
+\lambda_{7} \left(\Phi^{\dagger} \Xi\right)^2
\nonumber \\
&&
\hspace{-0.4cm}
+\lambda_{8}\left(\Phi^{\dagger} \Xi\right)\left(\Xi^{\dagger} \Phi\right)
+\frac{\lambda_{10}}{\sqrt{2}} \, \varepsilon_{ijk} \, \Phi_{i} \, \Lambda_{j} \, \Xi_{k} + \mbox{h. c.}
\; ,
\end{eqnarray}
and the Yukawa sector under $G_{331} \times \mathbb{Z}_{2}$ is given by
\begin{eqnarray}\label{LYZ2}
-{\cal L}_{Y}^{\mathbb{Z}_{2}} &=&
f_{ij}^{(e)} \, \overline{\psi}_{iL} \, \Lambda \, e_{jR}
+f_{ij}^{\prime(e)} \, \varepsilon_{i'j'k'} \, \left(\overline{\psi}_{iL}\right)_{i'}
\, \left(\psi_{jL}\right)_{j'}^{c}
\, \left(\Lambda^{\ast}\right)_{k'}
+f_{aj}^{(u)} \, \overline{q}_{aL} \, \Lambda^{\ast} \, u_{jR}
\nonumber \\
&&
\hspace{-0.4cm}
+ f_{1j}^{(d)} \, \overline{Q}_{L} \, \Lambda \, d_{jR}
+ f_{ij}^{(N)} \, \overline{\psi}_{iL} \, \Phi \, N_{jR}
+ f_{1i}^{(u)} \, \overline{Q}_{L} \, \Phi \, u_{iR}
+ f_{aj}^{(d)} \, \overline{q}_{aL} \, \Phi^{\ast} \, d_{jR}
\nonumber \\
&&
\hspace{-0.4cm}
+f_{44}^{(u)} \, \overline{Q}_{L} \, \Xi \, u_{4R}
+f_{a(a+2)}^{(d)} \, \overline{q}_{aL} \, \Xi^{\ast} \, d_{(a+2)R}
+M_{Rij} \, \overline{N^{c}}_{iR} \, N_{jR}
+\mbox{h. c.} \; .
\end{eqnarray}
The $\mathbb{Z}_{2}$ symmetry also allows to interpret the scalar triplet $\Xi$
as the responsible to break the gauge symmetry at the TeV scale and yield masses for heavy
gauge bosons. Furthermore, the new discrete symmetry also include the stability needed to the
new particle content of the model to be interpreted as DM candidate \cite{MonteroPRD2018,Cogollo}.
From now on, we consider in this paper the SSM mechanism ruled by the Higgs potential (\ref{HiggspotentialZ2}),
as well as, after the SSB takes place, the fermion masses are generated by Yukawa Lagrangian (\ref{LYZ2}).
The next section is dedicated to masses of the neutral and charged gauge bosons, in which we fix
conditions for the VEVs scales to defined the TeV scale. We also obtain the physical eigenstates and
the couplings of the fermions with the neutral gauge bosons.

\section{The masses of gauge bosons and neutral currents}
\label{sec3}

After the SSB mechanism takes place, the charged heavy bosons sector is given by
\begin{eqnarray}\label{LGaugemassesXB}
{\cal L}_{mass}^{WY}=m_{W}^2 \, W_{\mu}^{\, +}W^{\mu-}
+m_{Y}^2 \, Y_{\mu}^{\, +}Y^{\mu-} \; ,
\end{eqnarray}
%
%
%
%
where the masses of $W^{\pm}$
and $Y^{\pm}$ are, respectively, given by
\begin{eqnarray}
m_{W}=\frac{g_{L}}{2} \sqrt{v_{1}^{2}+v_{2}^{2}}
\hspace{0.4cm} \mbox{and} \hspace{0.4cm}
m_{Y}=\frac{g_{L}}{2}\sqrt{v_{3}^2+v_{2}^{2}} \; .
\end{eqnarray}
The couplings of the charged gauge fields $W^{\pm}$, $Y^{\pm}$ and neutral gauge field $V_{0}$ with the fermions of the model reads below :
\begin{eqnarray}\label{LintWXY}
-{\cal L}_{int}^{WV_{0}Y}&=&\frac{g_{L}}{\sqrt{2}} \, \overline{\nu}_{iL} \, \slash{\!\!\!W}^{+} \, e_{iL}
+\frac{g_{L}}{\sqrt{2}} \, \overline{u}_{iL} \, \, \slash{\!\!\!\!W}^{+} \, d_{iL} +
\nonumber \\
&&
\hspace{-1cm}
+\frac{g_{L}}{\sqrt{2}} \, \left( \, \overline{u}_{4L} \, \, \slash{\!\!\!Y}^{+} \, d_{1L}
+\overline{u}_{aL} \, \, \slash{\!\!\!Y}^{+} \, d_{(a+2)L}
+\overline{u}_{1L} \, \, \slash{\!\!\!V}_{0}  \, u_{4L}
+\overline{d}_{(a+2)L} \, \, \slash{\!\!\!V}_{0} \, d_{aL} \, \right)
+\mbox{h. c.} \; .
\end{eqnarray}
%
%
%
%
Note that the couplings of the quarks with $W_{\mu}^{\pm}$ are reproduced here like in the SM, and the charged gauge bosons
$Y_{\mu}^{\pm}$ mix the quarks of the SM with the new content of the 331 model, $u_{4L}$ , $d_{4L}$ and $d_{5L}$.
The sector of neutral gauge bosons has the content of three vector fields : $W_{3L\mu}$, $W_{8L\mu}$ and $B_{\mu}$.
After the SSB mechanism, the mass of neutral gauge bosons can be casted in the matrix form :
%
\begin{eqnarray}
{\cal L}_{mass}=
\frac{1}{2} \, \left(V_{\mu}\right)^{\!t} \, \eta^{\mu\nu} \, M^{2} \, V_{\nu} \; ,
\end{eqnarray}
where $(V_{\mu})^{t}=\left( \; W_{3L\mu} \; \; W_{8L\mu} \; \; B_{\mu} \; \right)$, $\eta^{\mu\nu}=\mbox{diag}\left(+1,-1,-1,-1\right)$
is the Minkowski metric,
%
%
and the mass matrix $M^{2}$ is given by
\begin{equation}\label{MatrixMassa}
M^{2}=
\left(
\begin{array}{ccc}
\frac{g_{L}^{2}}{4}\left(v_{1}^2+v_{2}^{2}\right) & \frac{g_{L}^2}{4\sqrt{3}}\left(v_{1}^{2}-v_{2}^{2} \right) & -\frac{g_{L}g_{X}}{6}\left(v_{1}^2+2v_{2}^{2} \right)
\\
\\
\frac{g_{L}^2}{4\sqrt{3}}\left(v_{1}^{2}-v_{2}^{2} \right) & \frac{g_{L}^2}{12}\left(v_{1}^2+v_{2}^{2}+4v_{3}^2\right) & \frac{g_{L}g_{X}}{6\sqrt{3}} \left(2v_{2}^{2}-v_{1}^2+2v_{3}^2\right)
\\
\\
-\frac{g_{L}g_{X}}{6}\left( v_{1}^2+2v_{2}^{2} \right) & \frac{g_{L}g_{X}}{6\sqrt{3}} \left(2v_{2}^{2}-v_{1}^2+2v_{3}^2\right) & \frac{g_{X}^2}{9}(v_{1}^2+4v_{2}^2+v_{3}^2)
\\
\end{array}
\right) \, .
\end{equation}
%
%
%

%
The diagonalization of mass matrix is carry out an $SO(3)$-orthogonal transformation
$V \; \longmapsto \; \tilde{V}_{\mu} = R^{t} \, V_{\mu} $, where $R$ is the most general orthogonal matrix of $SO(3)$.
%
The Lagrangian is written in the basis $\tilde{V}_{\mu}$ in which the diagonal mass matrix is
%
$M_{diag}^{2}=R^{t} \, M^{2} \, R=\mbox{diag}\left( \, m_{Z}^{\, 2} \, , \, m_{Z^{\prime}}^{\, 2} \, , \, 0 \, \right)$.
We identify the non-null elements of the diagonal as the masses of $Z$ and $Z^{\prime}$ in terms of VEVs scales and the coupling
constants
%
\begin{eqnarray}\label{massesZZ_{R}}
m_{Z} \simeq \frac{g_{L}}{2} \, \sqrt{v_{1}^2+v_{2}^2} \; \sqrt{\frac{3 g_{L}^2+4 g_{X}^2}{3 g_{L}^2+g_{X}^2}  }
\hspace{0.5cm} \mbox{and} \hspace{0.5cm}
m_{Z^{\prime}} \simeq \frac{g_{L} v_{3}}{\sqrt{3}} \, \sqrt{ 1 + \frac{g_{X}^{\, 2}}{3g_{L}^{2}}} \; ,
\end{eqnarray}
in which we have assumed $v_{3} \gg \left( \, v_{1} \, , \, v_{2} \, \right)$.
The null mass eigenstate is identified as the electromagnetic massless photon.
The $SO(3)$-matrix is parameterized by three rotation angles, {\it i. e.},
$\left\{ \, \theta_{1} \, , \, \theta_{2} \, , \, \theta_{3} \, \right\}$, in which $R=R_{1}(\theta_{1}) \, R_{2}(\theta_{2}) \, R_{3}(\theta_{3})$.
The mass basis of gauge fields is composite by the physical fields
$\left(\tilde{V}_{\mu}\right)^{t}=\left( \; Z_{\mu} \; \; Z_{\mu}^{\prime} \; \; A_{\mu} \; \right)$,
where the old gauge fields are written in terms of physical fields by the inverse transformation
$V_{\mu}=R \, \tilde{V}_{\mu}$. Using these transformations, the second rotation angle is identified
as the Weinberg angle $\theta_{2}=\theta_{W}$, in which $\sin^{2}\theta_{W} \simeq 0.23$.
The coupling constants $g_{L}$, $g_{X}$ and the mixing angles $\left\{ \; \theta_{1} \, , \, \theta_{W} \; \right\}$ obey the
electric charge parametrization
%
%
%
%
%
\begin{eqnarray}\label{chargee}
e=g_{L}\sin\theta_{W}
=g_{X} \, \cos\theta_{W} \, \sqrt{1-\frac{1}{3} \tan^2\theta_{W}}\; ,
\end{eqnarray}
where the fundamental charge is fixed by the fine structure constant
$e^{2}=4\pi/128\simeq0.09$, the $\theta_{1}$-angle is given by $\sin\theta_{1}=-\tan\theta_{W}/\sqrt{3}\simeq-0.31$
and $\cos\theta_{1}\simeq +0.95$.
What fixes the coupling constants at $g_{L}=0.64$ and $g_{X}=0.36$. Using these values,
we obtain the $W$ and $Z$ masses of SM : $m_{W}=80 \, \mbox{GeV}$ and $m_{Z}=91 \, \mbox{GeV}$.
The new gauge bosons $Z^{\prime}$ and $Y$ have the masses related by
\begin{eqnarray}\label{Masses}
m_{Z^{\prime}}=\frac{2 \, m_{Y}}{\sqrt{3-\tan^2\theta_{W}}} \; .
%
\end{eqnarray}
The lower limit obtained at the LHC experiment fixes the mass of charged gauge boson $Y$ at $m_{Y}=4.4$ TeV
\cite{CMS,ATLAS}. Therefore, relation (\ref{Masses}) fixes the $Z'$-mass at $m_{Z'}\simeq 5.36$ TeV.
These new masses defines a $v_{3}$-VEV scale at $v_{3}= 13.75$ TeV. The $\theta_{3}$ angle is proportional to ratio $m_{Z}^2/m_{Z'}^2$,
and using the previous numerical values, we obtain $\theta_{3} \simeq 0.0002882$ and we can assume $\cos\theta_{3}\simeq 1$ and $\sin\theta_{3}\simeq 0$ in the expressions from now on.
In the basis fields $Z_{\mu}$, $Z_{\mu}^{ \, \prime}$ and $A_{\mu}$ , we obtain the inverse transformations
\begin{eqnarray}\label{transfA0CGY}
W_{3L\mu} &=& \cos\theta_{W} \, Z_{\mu}
+ \sin\theta_{W} A_{\mu} \; ,
\nonumber \\
W_{8L\mu} &=&
\sqrt{1-\frac{1}{3} \tan^2\theta_{W}} \, Z^{\prime}_{\mu}
-\frac{\tan\theta_{W}}{\sqrt{3}} \left( - \sin\theta_{W} \, Z_{\mu}+\cos\theta_{W} \, A_{\mu} \right)  \; ,
\nonumber \\
B_{\mu} &=& \frac{\tan\theta_{W}}{\sqrt{3}} \, Z^{\prime}_{\mu}
+\sqrt{1-\frac{1}{3} \tan^2\theta_{W}} \left( - \sin\theta_{W} \, Z_{\mu}+\cos\theta_{W} \, A_{\mu} \right) \; .
\end{eqnarray}
The neutral gauge bosons $Z_{\mu}$, $Z_{\mu}^{\, \prime}$ and $A_{\mu}$ interact with any quiral fermion $\psi^{\, i}$ (left- or right-handed) of the model
via Lagrangian below
%
%
%
%
%
%
%
\begin{eqnarray}\label{LintAZZ}
{\cal L}_{NC}^{int} = e \, J_{\mu}^{\, \, (em)} A^{\mu}+\frac{g_{L}}{\cos\theta_{W}} \, J_{\mu}^{\, (Z)} \, Z^{\mu}
+ \frac{g_{X}}{\sqrt{3}\tan\theta_{W}} \, J_{\mu}^{\, (Z')} \, Z^{\prime \, \mu} \; ,
\end{eqnarray}
where $J_{\mu}^{\, \, (em)}=Q^{\, i}_{em} \, \overline{\psi^{\, i}}_{L(R)} \, \gamma_{\mu} \, \psi_{L(R)}^{\, i}$
is the usual EM current, $J_{\mu}^{\, (Z)} = J_{3L\, \mu}-\sin^2\theta_{W} J_{\mu}^{\, (em)}$ is the neutral current of $Z$ in the SM,
and the neutral current of $Z'$ is given by
%
%
\begin{eqnarray}
J_{\mu}^{\, (Z')} = \sqrt{3} \, J_{8L \, \mu}+\tan^2\theta_{W} \left( J_{\mu}^{\, \, (em)} - J_{3L \, \mu} \right)
\; . \; \; \; \;
\end{eqnarray}
The $Q^{\, i}_{em}$ is the electric charge from the equation (\ref{Qem}) for a $i$-fermion
in the table (\ref{table1}), similarly to EM current, the currents $J_{3L\, \mu}$ and $J_{8L\, \mu}$ are defined by
$J_{3L\, \mu}=I_{3L}^{\, i} \, \overline{\psi^{i}}_{L(R)} \, \gamma_{\mu} \, \psi_{L(R)}^{\, i}$
and $J_{8L\, \mu}=I_{8L}^{\, i} \, \overline{\psi^{i}}_{L(R)} \, \gamma_{\mu} \, \psi_{L(R)}^{\, i}$, respectively.
In these expressions, the index $i$ means a sum in all fermions of the model. All the values of isospin that we need to next
calculations are listed in the table (\ref{table1}). We can write the left- and right-handed fermions in terms of the
projectors, {\it i. e.}, $\psi_{L}^{\, i}=L \, \psi^{\, i}$ and $\psi_{R}^{\, i}=R \, \psi^{\, i}$, where
$L=({\uma}-\gamma_{5})/2$ and $R=({\uma}+\gamma_{5})/2$. Therefore, the $Z^{\prime}$ neutral current
can be written in terms of any non-quiral fermions $\psi^{\, i}$
\begin{eqnarray}\label{JZ_{R}Ja'}
J_{\mu}^{\, (Z')} = \frac{1}{2} \, c_{\psi_{L}^{\, i}} \, \overline{\psi^{\, i}} \, \gamma_{\mu} \, \left(1-\gamma_{5}\right) \, \psi^{\, i}
+\frac{1}{2} \, c_{\psi_{R}^{\, i}} \, \overline{\psi^{\, i}} \, \gamma_{\mu} \, \left(1+\gamma_{5}\right) \, \psi^{\, i} \; ,
\end{eqnarray}
where the coefficients $c_{\psi_{L}^{\, i}}$ and $c_{\psi_{R}^{\, i}}$ are defined by
\begin{eqnarray}
c_{\psi_{L}^{\, i}} = \sqrt{3} \, I_{8L}^{\, i}+\tan^2\theta_{W} \left( Q_{em}^{\, i} - I_{3L}^{\, i} \right)
\hspace{0.3cm} \mbox{and} \hspace{0.3cm}
c_{\psi_{R}^{\, i}} = \tan^2\theta_{W} \, Q_{em}^{\, i} \; .
\end{eqnarray}
%
%
%
%
%
%
%
%
Using the charges in the table (\ref{table1}), we list all the expressions of $c_{\psi_{L}^{\, i}}$ and $c_{\psi_{R}^{\, i}}$ :
\begin{eqnarray}\label{c}
c_{\nu_{L}} &=& 1-\frac{1}{2} \tan^2\theta_{W}
\hspace{0.2cm} , \hspace{0.2cm}
c_{e_{L}} = 1-\frac{1}{2} \tan^2\theta_{W}
\hspace{0.2cm} , \hspace{0.2cm}
\nonumber \\
c_{e_{R}} &=& -\tan^2\theta_{W}
\hspace{0.2cm} , \hspace{0.2cm}
c_{\nu_{L}^{c}} = -1
\hspace{0.2cm} , \hspace{0.2cm}
c_{N_{R}}=0 \; ,
\nonumber \\
c_{u_{1L}} &=& \frac{1}{2} + \frac{1}{6} \tan^2\theta_{W}
\hspace{0.3cm} , \hspace{0.3cm}
c_{u_{aL}} = -\frac{1}{2}+ \frac{1}{6} \tan^2\theta_{W}
\hspace{0.3cm} , \hspace{0.3cm}
\nonumber \\
c_{d_{1L}} &=& \frac{1}{2} + \frac{1}{6} \tan^2\theta_{W}
\hspace{0.3cm} , \hspace{0.3cm}
c_{d_{aL}} = -\frac{1}{2}+\frac{1}{6} \tan^2\theta_{W} \; ,
\nonumber \\
c_{u_{4L}} &=& -1 + \frac{2}{3} \tan^2\theta_{W}
\hspace{0.3cm} , \hspace{0.3cm}
c_{d_{(a+2)L}} = 1 - \frac{1}{3} \tan^2\theta_{W} \; ,
\nonumber \\
c_{u_{sR}} &=& +\frac{2}{3} \tan^2\theta_{W}
\hspace{0.3cm} , \hspace{0.3cm}
c_{d_{tR}}=-\frac{1}{3} \tan^2\theta_{W} \; .
\end{eqnarray}
All the interactions of SM fermions with the EM photon and the $Z$ gauge boson are reproduced in (\ref{LintAZZ}).
The content of conjugate left-handed neutrinos $\nu_{L}^{\, c}$ that emerge in the lepton's triplet interacts just with $Z'$.
The new content of exotic quarks $u_{4}$, $d_{4}$ and $d_{5}$ have electric charge $Q_{em}^{\, u_{4}}=+2/3$,
$Q_{em}^{\, d_{4}}=Q_{em}^{\, d_{5}}=-1/3$, and also can interact with $Z'$ via coefficients in (\ref{c}).
Since we know all the couplings of fermions with $Z'$, the $Z'$-decay width formula into any fermion $\psi^{\, i}$ is given by
\begin{equation}\label{DecayZprime}
\Gamma(Z' \rightarrow \overline{\psi^{\, i}} \, \psi^{\, i} )= N_{c}^{(\psi)} \, \frac{ m_{Z'} }{36\pi}
\frac{g_{X}^{\, 2}}{\tan^{2}\theta_{W}}
\sqrt{1-\frac{4m_{\psi}^{\, 2}}{m_{Z'}^{\, 2}}}
\left[ |c_{V}^{(\psi)}|^2 \left( 1 + \frac{2m_{\psi}^{\, 2}}{m_{Z'}^{\, 2}} \right) + |c_{A}^{(\psi)}|^2 \left( 1 - \frac{4m_{\psi}^{\, 2}}{m_{Z'}^{\, 2}} \right) \right] \; ,
\end{equation}
in which is constraint by the condition $m_{Z'} > 2 m_{\psi}$ for a fermion of mass $m_{\psi}$,
$N_{c}^{(\psi)}$ is the color number for a $\psi$-fermion : $N_{c}^{(\ell)}=1$ for charged leptons,
$N_{c}^{(\nu)}=1/2$ for neutrinos, and $N_{c}^{(Q)}=3$ for all quarks. The coefficients
$c_{V}^{(\psi)}=2 \, \left(c_{\psi^{\, i}_{L}} + c_{\psi^{\, i}_{R}}\right)$ and $c_{A}^{(\psi)}=2 \, \left(c_{\psi_{R}^{\, i}} - c_{\psi_{L}^{\, i}}\right)$ express the vector and axial components, respectively. Note that, if $\psi$ is any fermion $(f)$ in the SM, we can use the
approximation $m_{Z'} \gg m_{f}$ in (\ref{DecayZprime}) to obtain the simplified decay mode :
\begin{equation}\label{DecayZprimef}
\Gamma(Z' \rightarrow \overline{f} \, f ) \simeq N_{c}^{(f)} \, \frac{ m_{Z'} }{36\pi}
\frac{g_{X}^{\, 2}}{\tan^{2}\theta_{W}}\left( \, |c_{V}^{(f)}|^2 + |c_{A}^{(f)}|^2 \, \right) \; .
\end{equation}
Using the $Z'$ mass of $m_{Z'}=5.36$ TeV as an example, all the decay with into SM fermions have the values:
%
\begin{eqnarray}
\Gamma\left(Z' \rightarrow \bar{\nu} \, \nu \right) &\simeq&
90.41 \, \mbox{GeV}
\; , \;
\Gamma\left(Z' \rightarrow \bar{e} \, e \right) \simeq 183.46 \, \mbox{GeV} \; ,
\nonumber \\
\Gamma\left(Z' \rightarrow \bar{u} \, u \right) &\simeq& 637.09 \, \mbox{GeV}
\; , \;
\Gamma\left(Z' \rightarrow \bar{d} \, d \right) \simeq 634.44 \, \mbox{GeV} \; .
\end{eqnarray}
\section{The fermion eigenstates and the extra CP violation}
\label{sec4}
After the SSB takes place, the yukawa interactions in (\ref{LHiggs}) yield the leptons masses :
\begin{eqnarray}\label{Lmasse}
-{\cal L}_{mass}^{\, (e)} = \frac{v_{2}}{\sqrt{2}} \, f_{ij}^{(e)} \, \overline{e}_{iL} \, e_{jR}+\mbox{h. c.} \; .
\end{eqnarray}
The diagonalization of (\ref{Lmasse}) follows like in the SM. The $f_{ij}^{(e)}$ are the elements of the lepton's non-diagonal mass matrix.
This mass matrix can be diagonalized by the biunitary transformations
%
$e_{L} \longmapsto e'_{L}=U_{L} \, e_{L}$ and $e_{R} \longmapsto e'_{R}=U_{R} \, e_{R}$,
%
where $U_{L}$ and $U_{R}$ are unitary matrices, respectively. Thereby, using the spectral theorem, the lepton's mass matrix becomes
diagonal whose the real and positive eigenvalues are identified as the physical leptons masses :
%
$M_{D}^{(e)}=U_{L} \, \frac{v_{2} f^{(e)}}{\sqrt{2}} \, U_{R}^{\dagger}=\mbox{diag}(m_{e},m_{\mu},m_{\tau})$.
%
%
The prime basis $e'_{L}$ and $e'_{R}$ are the mass basis for leptons. From now on, the prime basis notation means the mass basis of physical fields for the fermions of the model.
The Yukawa sector (\ref{LHiggs}) also yields the Majorana masses for the left- and right-handed neutrinos :
\begin{eqnarray}\label{LpsichiM}
-{\cal L}_{mass}^{(N)} = M_{1ij}
\, \overline{\nu}_{iL} \, N_{jR}
+
M_{2ij} \, \overline{\nu_{iL}^{\, c}} \, \nu_{jL}
+M_{Rij} \, \overline{N_{iR}^{\, c}} \, N_{jR} + \mbox{h. c.} \; ,
\end{eqnarray}
%
%
%
%
%
where we have defined the elements $M_{1ij}=v_{1} \, f_{ij}^{(N)}/\sqrt{2}$ and $M_{2ij}=v_{2} \, f_{ij}^{\prime(e)}/\sqrt{2}$.
Note that $M_{2ij}$ are elements of a Dirac mass matrix. The mass sector in (\ref{LpsichiM}) can be casted in the matrix form
\begin{eqnarray}\label{LnuzetaMatrix}
-{\cal L}_{mass}^{(N)}=
\frac{1}{2} \, \overline{\chi_{i}^{c}} \, M_{ij}^{(N)} \, \chi_{j} + \mbox{h. c.} \; ,
\end{eqnarray}
in which the column vector of six elements is $\chi_{i}^{t}:=\left( \, \nu_{iL}^{c} \; \; N_{iR} \, \right)$,
and the $6 \times 6$ mass matrix $M^{(N)}$ is composite by four $3\times 3$ sub-matrices :
\begin{eqnarray}\label{MassMatrixM}
M^{(N)}=
\left(
\begin{array}{cc}
2M_{2} & \quad M_{1}
\\
\\
M_{1}^{t} & \quad 2 \, M_{R}  \\
\end{array}
\right) \; .
\end{eqnarray}
The matrix (\ref{MassMatrixM}) is diagonalized block-to-block through the unitary transformations
$
\chi \, \longmapsto \, \chi^{\prime} = U^{(N)} \, \chi
$,
where $U^{(N)}$ is a $6 \times 6$ unitary matrix. Thereby, the diagonal mass matrix is given by
$M_{D}^{(N)} = U^{(N)} \, M^{(N)} \, U^{(N)\dagger}=\mbox{diag}\left(
\mu_{-}^{(N)} \, , \, \mu_{+}^{(N)} \right)$, whose the diagonal sub-matrices are the eigenvalues of (\ref{MassMatrixM}) :
\begin{eqnarray}\label{MassesSeesaw}
\mu_{-}^{(N)} &=& \frac{1}{\sqrt{2}} \left[ \, M_{R} + M_{2} -\sqrt{ \left(M_{R}-M_{2}\right)^{2}
+ M_{1} \, M_{1}^{t} } \, \right] \, ,
\nonumber \\
\mu_{+}^{(N)} &=& \frac{1}{\sqrt{2}} \left[ \, M_{R} + M_{2} + \sqrt{ \left(M_{R}-M_{2} \right)^{2}
+M_{1} \, M_{1}^{t} } \, \right] \; .
\end{eqnarray}
Since we fix the RHNs masses around the $10$ TeV, we can use the approximation $M_{Rij} \gg M_{1ij}$
and $M_{Rij} \gg M_{2ij}$ for each element of the sub-matrices. Thereby, the neutrino masses in this seesaw
mechanism are the elements read below :
\begin{eqnarray}
m_{\nu_{L}ij} \simeq M_{2ij}-(M_{1} \, M_{R}^{-1} \, M_{1}^{\, t})_{ij}
\hspace{0.4cm} \mbox{and} \hspace{0.4cm}
m_{N_{R}ij} \simeq M_{Rij}
\; .
\end{eqnarray}
Therefore, this seesaw mechanism type II identifies the light LHNs with tiny mass elements bounded by the VEV $v_{2}$, and by
the coupling constants $f_{ij}^{\prime(e)}$. The RHN is the heavier particle of the model whose the elements $M_{Rij}$ are such that
$|M_{Rij}| \simeq 10$ TeV. The $3 \times 3$ matrices $m_{\nu_{L}}$ and $m_{N_{R}}$ can be diagonalized separately by other biunitary
transformation such that the diagonal matrices are the real and positive neutrinos masses. Using the measurement of transition probability,
we can bound $v_{2}$ and $f_{ij}^{\prime(e)}$ for the electron and muon neutrinos family. In this case, the normal hierarchy for neutrino masses of $\nu_{e}(1)$ and $\nu_{\mu}(2)$
is : $\Delta m_{12}^{\, 2}=|m_{2}^{2}-m_{1}^{2}|=(7.50 \pm 0.20) \times 10^{-5} \, \mbox{eV}^{2}$ \cite{BellPRL2005}.
Thus, the difference in the coupling constant $f^{\prime(e)}$ to the square times $v_{2}$ is constraint by $v_{2} \, \sqrt{|f_{2}^{\prime(e) \ast \, 2}-f_{1}^{\prime (e) \ast \, 2}|} \simeq 1.22 \times 10^{-2}$ eV.
The matrix $U^{(N)}$ can be written block to block in terms of four $3 \times3$ sub-matrices
\begin{eqnarray}
U^{(N)}=
\left(
\begin{array}{cc}
V_{1} & \quad \quad V_{2}
\\
\\
V_{3} & \quad \quad V_{4} \\
\end{array}
\right) \; ,
\end{eqnarray}
and substituting the transformation in the basis $\chi_{i}^{\prime}$, the coupling of leptons with the LHNs from
(\ref{LintWXY}) is
\begin{eqnarray}
{\cal L}_{\nu'e'W}^{int}=\frac{g_{L}}{\sqrt{2}} \, \overline{\nu^{\prime}}_{iL} \, (U_{PMNS})_{ij} \, \slash{\!\!\!W}^{+} \, e^{\prime}_{jL}
+\mbox{h. c.} \; ,
\end{eqnarray}
where $(U_{PMNS})_{ij}=U_{Lik}^{\ast}V_{1kj}$ are the elements of the Pontecorvo-Maki-Nakagawa-Sakata (PMNS) matrix.
In the standard form, the PMNS matrix is parameterized by three mixing angles $\left\{ \, \theta_{12} \, , \, \theta_{13} \, , \, \theta_{23} \, \right\}$, one Dirac phase $\delta_{D}$, and two Majorana CP phases $\delta_{1}$ and $\delta_{2}$:
%
%
$U_{PMNS}^{(L)}=U_{1}(\theta_{23}) \, U_{2}(\theta_{13},\delta_{D}) \, U_{3}(\theta_{12}) \, U_{4}(\delta_{1},\delta_{2})$,
%
where the $U_{i}$-matrices are given by
%
\begin{eqnarray}
U_{1}(\theta_{23}) &=&
\left(
\begin{array}{ccc}
1 & 0 & 0 \\
0 & \cos\theta_{23} & \sin\theta_{23} \\
0 & -\sin\theta_{23} & \cos\theta_{23} \\
\end{array}
\right)
\; , \;
U_{2}(\theta_{13},\delta_{D}) =
\left(
\begin{array}{ccc}
\cos\theta_{13} & 0 & e^{-i\delta_{D}} \sin\theta_{13} \\
0 & 1 & 0 \\
-e^{i\delta_{D}} \sin\theta_{13} & 0 & \cos\theta_{13} \\
\end{array}
\right) \; , \;
\nonumber \\
U_{3}(\theta_{12}) &=&
\left(
\begin{array}{ccc}
\cos\theta_{12} & \sin\theta_{12} & 0 \\
-\sin\theta_{12} & \cos\theta_{12} & 0 \\
0 & 0 & 1 \\
\end{array}
\right)
\; , \;
U_{4}(\delta_{1},\delta_{2}) =
\left(
\begin{array}{ccc}
1 & 0 & 0 \\
0 & e^{-i \delta_{1}} & 0 \\
0 & 0 & e^{-i \delta_{2}} \\
\end{array}
\right) \; .
\end{eqnarray}
The most recent constraints for these mixing angles are $\theta_{12}=33.82^{o}$, $\theta_{13}=8.61^{o}$,
$\theta_{32}=49.6^{o}$, and for the CP Dirac phase $\delta_{D}=215^{o}$, with $3\sigma$ range in the normal mass
ordering \cite{EstebanJHEP2019}.

The sector of quarks of the 331 model, including the new content $u_{4}$, $d_{4}$, and $d_{5}$, has the massive terms :
\begin{equation}\label{LmassQ}
-{\cal L}_{mass}^{(Q)} =
M_{ij}^{(u)} \, \overline{u}_{iL} \, u_{jR}
+M_{ij}^{(d)} \, \overline{d}_{iL} \, d_{jR}
+\frac{v_{3}}{\sqrt{2}} \, f_{44}^{(u)} \, \overline{u}_{4L} \, u_{4R}
+M_{(a+2)(b+2)}^{(d_{4}-d_{5})} \, \overline{d}_{(a+2)L} \, d_{(b+2)R}
+\mbox{h. c.}  \; ,
\end{equation}
where the $M_{ij}^{(u)}$ and $M_{ij}^{(d)}$ are the matrices
\begin{eqnarray}\label{MuMd}
M^{(u)}=
\frac{1}{\sqrt{2}}
\left(
\begin{array}{ccc}
f_{11}^{(u)}v_{2} & f_{12}^{(u)}v_{2} & f_{13}^{(u)}v_{2} \\
f_{21}^{(u)}v_{1} & f_{22}^{(u)}v_{1} & f_{23}^{(u)}v_{1} \\
f_{31}^{(u)}v_{1} & f_{32}^{(u)}v_{1} & f_{33}^{(u)}v_{1} \\
\end{array}
\right)
\; , \;
M^{(d)}=
\frac{1}{\sqrt{2}}
\left(
\begin{array}{ccc}
f_{11}^{(d)}v_{1} & f_{12}^{(d)}v_{1} & f_{13}^{(d)}v_{1} \\
f_{21}^{(d)}v_{2} & f_{22}^{(d)}v_{2} & f_{23}^{(d)}v_{2} \\
f_{31}^{(d)}v_{2} & f_{32}^{(d)}v_{2} & f_{33}^{(d)}v_{2} \\
\end{array}
\right) \; .
\end{eqnarray}
The hierarchy condition $M_{ij}^{(q)} \leq M_{(i+1)j}^{(q)}$, where $q=u,d$, must be satisfied to insurance
the suppression of flavour changing neutral currents (FCNCs) by the Froggatt-Nielsen mechanism \cite{Huitu}.
The matrix elements $M_{(a+2)(b+2)}^{(d_{4}-d_{5})}$ mixes the content of exotic quarks $d_{4}-d_{5}$
\begin{eqnarray}\label{Mab}
M^{(d_{4}-d_{5})}=\frac{v_{3}}{\sqrt{2}}
\left(
\begin{array}{cc}
f_{44}^{(d)} \quad & f_{45}^{(d)}
\\
\\
f_{54}^{(d)} \quad & f_{55}^{(d)} \\
\end{array}
\right) \; .
\end{eqnarray}
In (\ref{LmassQ}), since $f_{44}^{(u)}$ is a complex constant, so we write $f_{44}^{(u)}=|f_{44}^{(u)}| \, e^{i\delta_{u_4}}$
in which $\delta_{u_4}$ is a real phase. Since the mass of $u_{4}$ must be a physical quantity (real and positive), it is identified as $m_{u_{4}}=v_{3}\,|f_{44}^{(u)}|/\sqrt{2}$, where the phase $\delta_{u_4}$ is absorbed into the field $u_{4}$.
Using the VEV scale $v_{3}=13.75$ TeV constraint in the section \ref{sec3}, the mass of $u_{4}$ fermion is upper bounded by $m_{u_{4}} < 9.72$ TeV.
The matrices in (\ref{MuMd}) can be diagonalized by biunitary transformations, like in the SM : $u_{L(R)} \mapsto u'_{L(R)}=V_{L(R)}^{(u)} \, u_{L(R)}$
and $d_{L(R)} \mapsto d'_{L(R)}=V_{L(R)}^{(d)} \, d_{L(R)}$, where $V_{L(R)}^{(u)}$ and $V_{L(R)}^{(d)}$ are $3 \times 3$
unitary matrices, and the resultant diagonal matrix yields the SM quarks $M_{D}^{(u)}= V_{L}^{\, (u)} \, M^{(u)} \, V_{R}^{\, (u)\dagger}
=\mbox{diag}\left( \, m_{u} \, , \, m_{c} \, , \, m_{t} \, \right)
$ and $M_{D}^{(d)}= V_{L}^{\, (d)} \, M^{(d)} \, V_{R}^{\, (d)\dagger} =\mbox{diag}\left( \, m_{d} \, , \, m_{s} \, , \, m_{b} \, \right)$, respectively.
The matrix (\ref{Mab}) also can be diagonalized by another biunitary transformation, {\it i. e.},
$\chi_{L(R)} \mapsto \chi'_{L(R)}= V_{L(R)} \, \chi_{L(R)}$, where $\chi_{L(R)}^{t}=\left( \, d_{4L(R)} \; \; d_{5L(R)} \, \right)$
and $V_{L(R)}$ are $2 \times 2$ unitary matrices. The diagonal matrix is $M_{D}^{(4-5)}=V_{L} \, M^{(d_{4}-d_{5})} \, V_{R}^{\, \dagger}=\mbox{diag}\left( \, m_{d_{4}} \, , \, m_{d_{5}} \, \right)$ whose the elements are the real and positive eigenvalues of (\ref{Mab}):
\begin{eqnarray}
m_{d_{4}} &=& \frac{v_{3}}{2\sqrt{2}} \left[ \, |f_{44}^{(d)}|+|f_{55}^{(d)}| - \sqrt{ \left( |f_{44}^{(d)}| - |f_{55}^{(d)}| \right)^2+4\,|f_{45}^{(d)}||f_{54}^{(d)}| }   \, \right] \; ,
\nonumber \\
m_{d_{5}} &=& \frac{v_{3}}{2\sqrt{2}} \left[ \, |f_{44}^{(d)}|+|f_{55}^{(d)}| + \sqrt{ \left( \, |f_{44}^{(d)}| - |f_{55}^{(d)}| \right)^2+4\,|f_{45}^{(d)}||f_{54}^{(d)}| }  \, \right] \; ,
\end{eqnarray}
that is the content of mass for the exotic fermions $d_{4}$ and $d_{5}$. The mass $m_{d_{4}}$ is positive if the
couplings satisfy the condition $|f_{44}^{(d)}||f_{55}^{(d)}|>|f_{45}^{(d)}||f_{54}^{(d)}|$. Note that we have chosen the eigenvalues such that
$m_{d_{5}} > m_{d_{4}}$. Although the masses are fixed by the VEV scale $v_{3}$, the exotic quark $d_{4}$ must be the next generation
in relation to $d_{5}$, then we have considered, in principle, $d_{5}$ heavier than $d_{4}$. Using all the transformations in the
prime basis, the full diagonal Lagrangian for quark masses is
\begin{eqnarray}
-{\cal L}_{mass}^{(Q)}=\overline{u'}_{iL} \, M_{Dij}^{(u)} \, u'_{jR}
+\overline{d'}_{iL} \, M_{Dij}^{(d)} \, d'_{jR}
+m_{u_{4}} \, \overline{u}_{4} \, u_{4}+ \overline{\chi'}_{(a+2)L} \, M_{D(a+2)(b+2)}^{(d_{4}-d_{5})} \, \chi'_{(b+2)R}
\; .
\end{eqnarray}
%
%
%
%

%
Using the prime basis for the quarks fields, we obtain the interactions of the quarks
with the charged gauge bosons $W^{\pm}$, $Y^{\pm}$, and the neutral gauge bosons $V_{0}$ and $\bar{V}_{0}$
\begin{eqnarray}\label{LintQWW_{R}}
-{\cal L}_{Q-WYV_0}^{\, int} &=& \frac{g_{L}}{\sqrt{2}} \, \overline{u'}_{iL} \, V_{CKM ij} \, \, \slash{\!\!\!\!W}^{+} \,  d'_{jL}
+\frac{g_{L}}{\sqrt{2}} \, \overline{u}_{4L} \, \, \slash{\!\!\!Y}^{+} \, V^{(d)\dagger}_{L1i} \, d_{iL}^{\prime}
\nonumber \\
&&
\hspace{-0.5cm}
+\frac{g_{L}}{\sqrt{2}} \, \overline{u'}_{jL} \, V_{j(a+2)}^{(ud)} \, \, \slash{\!\!\!Y}^{+} \, d'_{(a+2)L}
+\frac{g_{L}}{\sqrt{2}} \, \overline{u'}_{iL} \, V^{(u)}_{L1i} \, \, \slash{\!\!\!V}_{0} \, u_{4L}
\nonumber \\
&&
\hspace{-0.5cm}
+\frac{g_{L}}{\sqrt{2}} \, \overline{d'}_{(a+2)L} \, V_{(a+2)j}^{(d)} \, \, \slash{\!\!\!V}_{0} \,  d'_{jL}
+\mbox{h. c.} \; ,
\end{eqnarray}
where $V_{CKM ij}$ are the elements of the usual Cabibbo-Kobayashi-Maskawa (CKM) matrix $V_{CKM}=V_{L}^{(u)}\,V_{L}^{(d)\dagger}$.
We use the standard parametrization for CKM matrix with three mixing angles $\{ \, \beta_{c} \, , \, \beta_{13} \, , \, \beta_{23} \, \}$ and one phase $\delta$ :
$
V_{CKM}=V_{1}(\beta_{23}) \, V_{2}(\beta_{13},\delta) \, V_{3}(\beta_{c})
$ ,
where $V_{i} \, (i=1,2,3)$ are rotation matrices, and $\beta_{c}=13.04^{o}$ is the Cabibbo angle, and the others two angles
are given by $\beta_{13}=0.201^{o}$, $\beta_{23}=2.38^{o}$, and the phase is $\delta=1.20$ rd.
The elements $V^{(d)\dagger}_{L1i}$ mix the quarks $(d,s,b)$ of the SM with the exotic quark $u_{4L}$ mediated by
the charged gauge boson $Y$, and $V^{(ud)}_{j(a+2)}$
mixes the quarks $(u,c,t)$ with the exotic content $d_{4L}$ and $d_{5L}$ also mediated by $Y$, respectively.
For the case of $V^{(d)\dagger}_{L1i}$, the most general $3 \times 3$ unitary matrix can be parameterized by three angles more six phases.
Five of these phases can be redefined in the fermion fields resting one phase. Thereby, the $3 \times 3$ matrix $V^{(d)\dagger}_{L}$ has
one phase that violate the CP symmetry in the 331 model. If this matrix is parameterized similarly to the CKM matrix, the coupling of
$u_{4L}$ with the bottom quark is
\begin{eqnarray}\label{LintCP}
-{\cal L}_{CP}^{\, int} &=&
\frac{g_{L}}{\sqrt{2}} \, \sin\gamma_{13} \, e^{i\delta_{b}} \, \overline{u}_{4L}  \, \, \slash{\!\!\!Y}^{+} \, b_{L}^{\prime}
+ \mbox{h. c.} \; ,
\end{eqnarray}
where
$\gamma_{13}$ is a mixing angle,
$\delta_{b}$ is a phase that violates the CP symmetry.
The magnitude of the coupling $g_{L}\sin\gamma_{13}$ in (\ref{LintCP}) must be weaker in relation to the CKM matrix.
%
%
%
%
%
%

%
These couplings mediated by the charged boson $Y$ contains a new the phenomenology of particle physics
beyond the SM at the TeV scale. As example,
$Y^{+}$ can decay into pairs $\overline{u}_{4} \, d'_{i}$,
and also into $\overline{u'}_{i} \, d'_{a+2}$.
The masses of the exotic quarks and charged gauge boson must satisfy kinetic conditions to allow the decay modes.
%

%

%
%

%

%

%
%
%
%

%

%
\section{The Higgs masses and the physical scalar fields}
\label{sec5}
In this section, we return to the sector of scalar fields with the ${\mathbb Z}_{2}$ potential (\ref{HiggspotentialZ2})
after the SSB mechanism. We start with the three scalar fields $\Phi$, $\Lambda$ and $\Xi$ parameterized such that
\begin{eqnarray}
\phi_{1}^{0}=\frac{v_{1}+h_{1}+iz_{1}}{\sqrt{2}}
\hspace{0.4cm} , \hspace{0.4cm}
\lambda^{0}=\frac{v_{2}+h_{2}+iz_{2}}{\sqrt{2}}
\hspace{0.4cm} , \hspace{0.4cm}
\xi_{2}^{0}=\frac{v_{3}+h_{3}+iz_{3}}{\sqrt{2}} \; , \; \;
\end{eqnarray}
%
%
in which $h_{1}$, $h_{2}$ and $h_{3}$ are three scalar fields mixed by the Higgs potential (\ref{HiggspotentialZ2}), and $z_{1}$,
$z_{2}$ and $z_{3}$ are the imaginary part of the scalar fields. The square matrix mass of $h_{1}$, $h_{2}$ and $h_{3}$ is :
\begin{eqnarray}\label{MH2}
M_{H}^{\, 2}=
\left(
\begin{array}{ccc}
2\lambda_{1}v_{1}^2+\frac{\lambda_{10}}{2}\frac{v_{2}v_{3}}{v_{1}} & -\frac{\lambda_{10}v_{3}}{2}+v_{1}v_{2}\lambda_{4} & -\frac{\lambda_{10}v_{1}}{2}+v_{2}v_{3}\lambda_{6}
\\
\\
-\frac{\lambda_{10}v_{3}}{2}+v_{1}v_{2}\lambda_{4} & 2\lambda_{2}v_{2}^2+\frac{\lambda_{10}}{2}\frac{v_{1}v_{3}}{v_{2}} & -\frac{\lambda_{10}v_{2}}{2}+v_{1}v_{3}\lambda_{5}
\\
\\
-\frac{\lambda_{10}v_{1}}{2}+v_{2}v_{3}\lambda_{6}  & -\frac{\lambda_{10}v_{2}}{2}+v_{1}v_{3}\lambda_{5}  & 2\lambda_{3}v_{3}^2+\frac{\lambda_{10}}{2}\frac{v_{1}v_{2}}{v_{3}} \\
\end{array}
\right) \; .
\end{eqnarray}
The matrix (\ref{MH2}) is diagonalized by a $SO(3)$ matrix. We denote the physical scalar fields in this sub-sector as $H_{1}$, $H_{2}$ and $H_{3}$
in which $H_{1}$ is the SM Higgs with mass of $125$ GeV. The eigenvalues of (\ref{MH2}) are simplified when we apply
the approximation $v_{3} \gg \left( \, v_{1} \, , \, v_{2} \, \right)$
\begin{eqnarray}\label{massesHi}
m_{H_{1}} \simeq \sqrt{2 \, \frac{\lambda_{1}v_{1}^4+\lambda_{4} \, v_{1}^2 \, v_{2}^2+\lambda_{2}v_{2}^4}{v_{1}^2+v_{2}^2}}
\hspace{0.2cm} , \hspace{0.2cm}
m_{H_{2}} \simeq \sqrt{ \frac{\lambda_{10}v_{3}}{2} \frac{v_{1}^2+v_{2}^2}{v_{1} \, v_{2}} }
\hspace{0.2cm} , \hspace{0.2cm}
m_{H_{3}} \simeq \sqrt{2 \, \lambda_{3} \, v_{3}^{2}} \; .
\end{eqnarray}
The real masses fix the positive couplings $\lambda_{3} > 0$ and $\lambda_{10} > 0$, in which we choose $\lambda_{10} =k \, v_{3}$
with $k>0$ of order 1. Using this approximation of the VEVs scales, the physical fields are obtained by the transformations
\begin{eqnarray}
\left(
\begin{array}{c}
h_{1}
\\
h_{2} \\
\end{array}
\right)
&\simeq&\frac{1}{\sqrt{v_{1}^2+v_{2}^2}}
\left(
\begin{array}{cc}
v_{1} \quad & v_{2}
\\
v_{2} \quad & -v_{1} \\
\end{array}
\right)\!
\left(
\begin{array}{c}
H_{1}
\\
H_{2} \\
\end{array}
\right)
\hspace{0.5cm} \mbox{and} \hspace{0.5cm}
 h_{3} \simeq H_{3} \; .
\end{eqnarray}
The squared mass matrix associated with the imaginary part $z_{1}$, $z_{2}$ and $z_{3}$ is
\begin{eqnarray}
M_{z}^{2}=
\left(
\begin{array}{ccc}
\frac{v_{2}v_{3}}{v_{1}} & v_{3} & v_{1} \\
v_{3} & \frac{v_{1}v_{3}}{v_{2}} & v_{2} \\
v_{1} & v_{2} & \frac{v_{1}v_{2}}{v_{3}} \\
\end{array}
\right) \; ,
\end{eqnarray}
whose the eigenvalues are given by
\begin{eqnarray}
m_{h_{0}}=\sqrt{\frac{\lambda_{10}}{2} \, \frac{v_{1}^2 \, v_{2}^2+v_{3}^{2} \, (v_{1}^2+v_{2}^2)}{v_{1} \, v_{2} \, v_{3}} }
\hspace{0.4cm} , \hspace{0.4cm}
m_{G_{1}}=m_{G_{2}}=0 \; ,
\end{eqnarray}
and the correspondent transformations are
\begin{eqnarray}
\left(
\begin{array}{c}
z_{2}
\\
z_{1} \\
\end{array}
\right)
&\simeq&\frac{1}{\sqrt{v_{1}^2+v_{2}^2}}
\left(
\begin{array}{cc}
v_{1} \quad & -v_{2}
\\
v_{2} \quad & v_{1} \\
\end{array}
\right)\!
\left(
\begin{array}{c}
h_{0}
\\
G_{1} \\
\end{array}
\right)
\hspace{0.5cm} \mbox{and} \hspace{0.5cm}
z_{3} \simeq G_{2} \; .
\end{eqnarray}
The $G_{1}$, $G_{2}$ are Goldstone bosons eaten by $Z$ and $Z^{\prime}$, respectively. The scalar field $h_{0}$ is the CP-odd Higgs boson.
The sector of charged fields is set by the matrices
\begin{eqnarray}\label{Mcharged}
M_{1}^{2}&=&\frac{\lambda_{10}v_{3}}{2}
\left(
\begin{array}{cc}
v_{2}/v_{1} \quad & -1
\\
-1 \quad & v_{1}/v_{2} \\
\end{array}
\right)
\hspace{0.1cm} , \hspace{0.1cm}
M_{2}^{2}=\frac{\lambda_{10}v_{1}}{2}
\left(
\begin{array}{cc}
v_{3}/v_{2} \quad & -1
\\
-1 \quad & v_{2}/v_{3} \\
\end{array}
\right)
\hspace{0.1cm} , \hspace{0.1cm}
\nonumber \\
M_{3}^{2}&=&\frac{\lambda_{10}v_{2}}{2}
\left(
\begin{array}{cc}
v_{3}/v_{1} \quad & -1
\\
-1 \quad & v_{1}/v_{3} \\
\end{array}
\right) \; .
\end{eqnarray}
Each matrix in (\ref{Mcharged}) is diagonalized, respectively, by the transformations
\begin{eqnarray}
\left(
\begin{array}{c}
\phi^{+}
\\
\lambda_{1}^{+} \\
\end{array}
\right)
&=&\frac{1}{\sqrt{v_{1}^2+v_{2}^2}}
\left(
\begin{array}{cc}
-v_{1} \quad & v_{2}
\\
v_{2} \quad & v_{1} \\
\end{array}
\right)\!
\left(
\begin{array}{c}
G_{1}^{+}
\\
H_{1}^{+} \\
\end{array}
\right) \; ,
\nonumber \\
\left(
\begin{array}{c}
\lambda_{2}^{+}
\\
\xi^{+} \\
\end{array}
\right)
&=&\frac{1}{\sqrt{v_{2}^2+v_{3}^2}}
\left(
\begin{array}{cc}
-v_{2} \quad & v_{3}
\\
v_{3} \quad & v_{2} \\
\end{array}
\right)\!
\left(
\begin{array}{c}
G_{2}^{+}
\\
H_{2}^{+} \\
\end{array}
\right) \; ,
\nonumber \\
\left(
\begin{array}{c}
\phi_{2}^{0}
\\
\xi_{1}^{\ast} \\
\end{array}
\right)
&=&\frac{1}{\sqrt{v_{1}^2+v_{3}^2}}
\left(
\begin{array}{cc}
-v_{1} \quad & v_{3}
\\
v_{3} \quad & v_{1} \\
\end{array}
\right)\!
\left(
\begin{array}{c}
G_{3}
\\
\chi \\
\end{array}
\right) \; ,
\end{eqnarray}
in which the eigenvalues are determinate by the masses
\begin{eqnarray}\label{massesscalarcharged}
m_{H_{1}^{\pm}} &\simeq& \sqrt{ \frac{\lambda_{10} v_{3}}{2} \, \frac{v_{1}^2+v_{2}^2}{v_{1}v_{2} } }
\hspace{0.3cm} , \hspace{0.3cm}
m_{G_{1}^{\pm}}=0
\hspace{0.3cm} , \hspace{0.3cm}
m_{H_{2}^{\pm}} \simeq \sqrt{ \frac{\lambda_{10} v_{1}}{2} \, \frac{v_{2}^2+v_{3}^2}{v_{2}v_{3} } }
\hspace{0.3cm} , \hspace{0.3cm}
\nonumber \\
m_{G_{2}^{\pm}}&=&0
\hspace{0.3cm} , \hspace{0.3cm}
m_{\chi} \simeq \sqrt{ \frac{\lambda_{10} v_{2}}{2} \, \frac{v_{1}^2+v_{3}^2}{v_{1}v_{3} } }
\hspace{0.3cm} , \hspace{0.3cm}
m_{G_{3}}=0 \; .
\end{eqnarray}
%
The fields $G_{1}^{\pm}$, $G_{2}^{\pm}$ and $G_{3}$ are Goldstone bosons adsorbed by the gauge transformations
of $W^{\pm}$, $X^{\pm}$ and $Y^{\pm}$, respectively. Note that, from (\ref{massesHi}) and (\ref{massesscalarcharged}),
$m_{H_{2}} \simeq m_{H_{1}^{\pm}}$, and using the condition $v_{3} \gg \left( \, v_{1} \, , \, v_{2} \, \right)$, the masses of
$H_{2}^{\pm}$ and $\chi$ satisfy to the relation $m_{H_{2}^{\pm}}/m_{\chi}\simeq v_{1}/v_{2}$.
If we parameterize the VEVs as $v_{1}=v \sin\alpha$ and $v_{2}=v \cos\alpha$, the derivative of the
potential (\ref{HiggspotentialZ2}) in relation to $\alpha$ angle yields the extremum condition
\begin{equation}\label{Eqalpha}
\left[ 2v_{3}^2(\lambda_{5}-\lambda_{6}+\lambda_{7}+\lambda_{8})+2(\mu_{1}^2-\mu_{2}^2)
+v^2(\lambda_{4}-2\lambda_{1}\sin^2\alpha-2\lambda_{2}\cos^2\alpha)+v_{3}^2 \cot(2\alpha) \right] \sin(2\alpha)=0 \; .
\end{equation}
One solution for the equation (\ref{Eqalpha}) is $\alpha=\pi/4$, with $v_{3}\simeq\sqrt{(\mu_{2}^2-\mu_{1}^{2})/(\lambda_{5}+\lambda_{7}+\lambda_{8}-\lambda_{6})}$, where the
parameters must satisfy the inequalities $\mu_{2}>\mu_{1}$ and $\lambda_{5}+\lambda_{7}+\lambda_{8}>\lambda_{6}$, or
$\mu_{2}<\mu_{1}$ and $\lambda_{5}+\lambda_{7}+\lambda_{8}<\lambda_{6}$. In this solution, the VEVs $\left( \, v_{1} \, , \, v_{2} \, \right)$
are given by $v_{1}=v_{2}\simeq 173.94$ GeV, and the masses of the scalars
are :
\begin{eqnarray}
m_{H_{1}} &=& \sqrt{ \left(\lambda_1+\lambda_2+\lambda_4\right) \frac{v^2}{2} } =125 \, \mbox{GeV} \; ,
\nonumber \\
m_{H_{1}^{\pm}} &=& m_{H_{2}} \simeq m_{h_{0}}= \sqrt{2} \, m_{H_{2}^{\pm}} = \sqrt{2} \, m_{\chi}  \simeq \sqrt{ k } \, v_3 \; ,
\end{eqnarray}
where $\lambda_1+\lambda_2+\lambda_4>0$.
%

%
In the physical fields basis, the coupling of $Z^{\prime}$ with SM $Z$ boson and Higgs $H_{1}$ is given by
%
%
\begin{eqnarray}\label{LintZH1Zp}
{\cal L}_{ZH_{1}Z'}^{\, int} &=& - \frac{g_{X}}{\sqrt{3}} \, \tan\theta_{W} \, m_{Z} \, H_{1} \, Z^{\mu} Z_{\mu}^{\prime}
\; ,
\end{eqnarray}
and the couplings of $Z^{\prime}$ with the scalars $H_{1}^{\pm}$, $H_{2}^{\pm}$ and $\chi$ are listed below :
\begin{eqnarray}
{\cal L}_{Z'H_{1}^{+}H_{1}^{-}}^{\, int} &\simeq& \frac{i \, g_{X}}{2\sqrt{3}} \, \frac{3+5\cos(2\theta_{W})}{\sin(2\theta_{W})} \left( \, H_{1}^{+} \, \partial^{\mu}H_{1}^{-}-H_{1}^{-} \, \partial^{\mu}H_{1}^{+} \, \right) Z_{\mu}^{\prime} \; ,
\nonumber \\
{\cal L}_{Z'H_{2}^{+}H_{2}^{-}}^{\, int} &\simeq& \frac{-i \, g_{X}}{\sqrt{3}} \, \frac{3+5\cos(2\theta_{W})}{\sin(2\theta_{W})} \left( \, H_{2}^{+} \, \partial^{\mu}H_{2}^{-}-H_{2}^{-} \, \partial^{\mu}H_{2}^{+} \, \right) Z_{\mu}^{\prime} \; ,
\nonumber \\
{\cal L}_{Z'\chi\chi^{\ast}}^{\, int} &\simeq& \frac{i \, g_{X}}{\sqrt{3} \, \tan\theta_{W}}  \, \left(\chi^{\ast} \, \partial^{\mu}\chi
-\chi \, \partial^{\mu}\chi^{\ast} \right) Z_{\mu}^{\prime}  \; .
\end{eqnarray}
%
%
%
%
%
The coupling (\ref{LintZH1Zp}) is associated with the $Z'$ decay into $Z$ gauge boson and the SM Higgs
$Z^{\prime} \rightarrow Z \, H_{1}$ that yields a contribution to the full width $Z^{\prime}$ decay at the SM scale.
The decay width of $Z^{\prime} \rightarrow Z \, H_{1}$ is read below
\begin{eqnarray}
\Gamma(Z^{\prime} \rightarrow Z \, H_{1}) &=& \frac{g_{X}^2}{192\pi} \, \sin^2\theta_{W} \,
m_{Z^{\prime}} \left[ x_{H_{1}}^4-2x_{H_{1}}^2(x_{Z}^2+1)+x_{Z}^2(10+x_{Z}^2)+1 \right]
\nonumber \\
&&
\times \sqrt{x_{H_{1}}^4-2x_{H_{1}}^2(x_{Z}^2+1)+(x_{Z}^2-1)^2}  \; ,
\end{eqnarray}
where we have defined $x_{H_{1}}=m_{H_{1}}/m_{Z'}$ and $x_{Z}=m_{Z}/m_{Z'}$. Using $m_{Z^{\prime}}=5.36$ TeV, and the masses of $H_{1}$ (125 GeV)
and $Z$ (91 GeV) in the SM, we obtain the result
\begin{eqnarray}
\Gamma(Z^{\prime} \rightarrow Z \, H_{1}) \simeq 0.0588 \, \mbox{GeV} \; .
\end{eqnarray}
The new content of physical scalar fields is rich at the TeV scale. Two charged fields $H_{1}^{\pm},H_{2}^{\pm}$,
two real scalar fields $H_{2},H_{3}$, and the complex scalar $\chi$ are fixed with masses at the TEV scale. The discrete symmetry
$\mathbb{Z}_{2}$ stabilizes the $\chi$ scalar field. It will be the candidate to the DM content of the model. Since it interacts
with $Z^{\prime}$, the couplings of $Z^{\prime}$ with the SM fermions make possible the annihilation processes needed to DM detection.
Therefore, the 331 model is an example of scalar DM content with the spin-1 portal set by the $Z^{\prime}$ gauge boson.
\section{The relic density and direct detection}
\label{sec6}
\begin{figure}[h]
\centering
\includegraphics[width=0.80\textwidth]{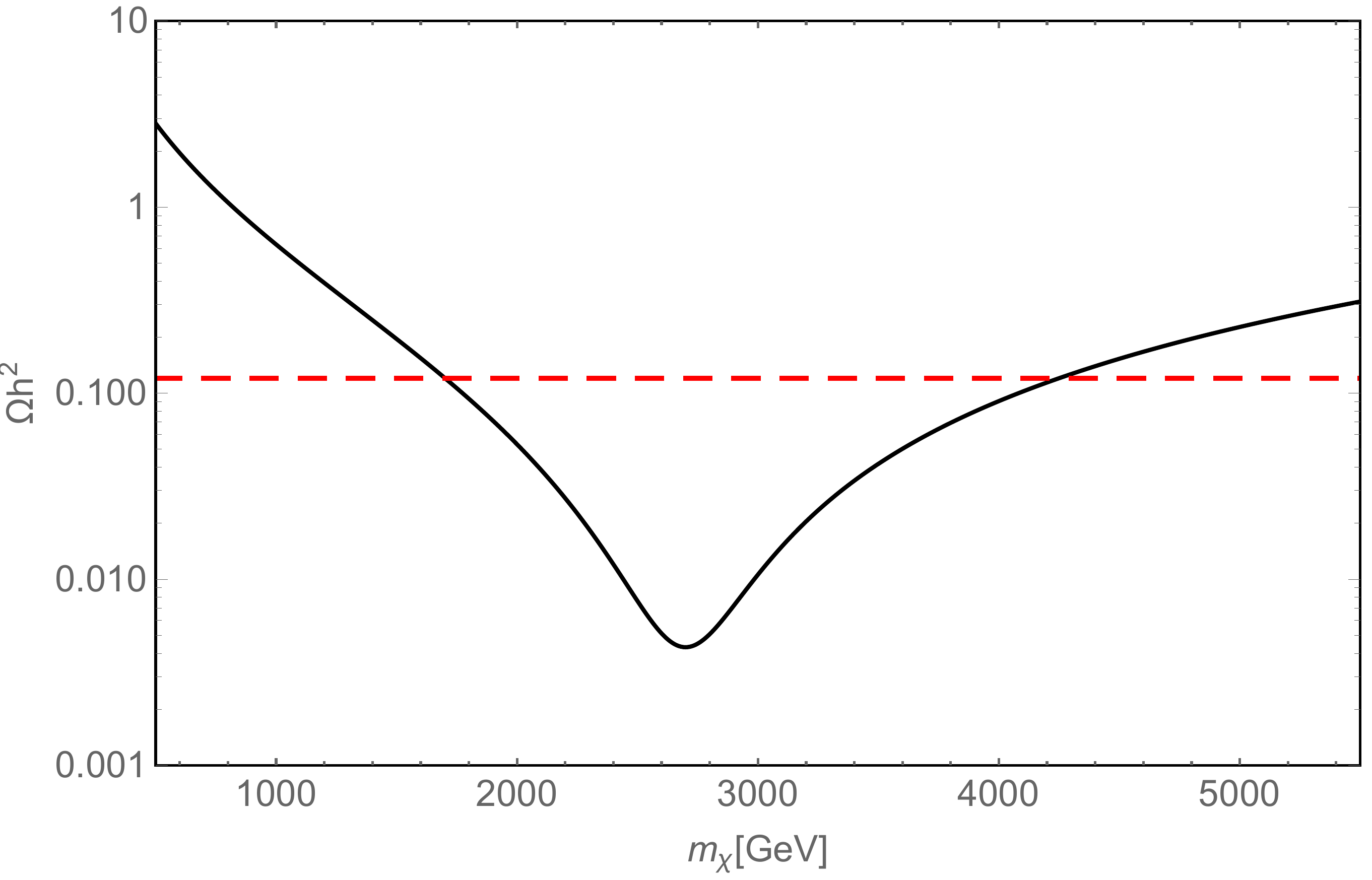}
\caption{The relic density as function of DM mass $m_{\chi}$. The $Z'$ mass is chosen at $5.36$ TeV
and $g_{X}=0.36$. The red dashed line means the observed relic density at $0.12$.}
\label{relic331}
\end{figure}

The scalar field $(\chi)$ is our candidate to DM content in this example of 331 model.
The ${\mathbb Z}_{2}$ discrete symmetry does the scalar field $\chi$ stable. For the relic density calculus, we use the
$Z^{\prime}$ mass constraint at $5.36$ TeV and the mass $m_{\chi}$ is a free parameter. The $Z'$ gauge boson works as the
DM portal to link the DM content with the SM fermions through annihilation processes. Two annihilation processes are important for the DM analysis : (i) The annihilation of DM-anti-DM scalar pair into SM fermion-anti-fermion pair via $Z'$ portal, through the $s$-channel by the process $\chi^{\ast} \, \chi \rightarrow Z^{\prime} \rightarrow \bar{f} \, f$, and (ii) The annihilation of DM-anti-DM scalar pair into $Z^{\prime}$ pair in the final state, {\it i. e.}, $\chi^{\ast} \, \chi \rightarrow Z^{\prime} Z^{\prime}$.
We start out the DM analysis with the Boltzmann equation :
\begin{eqnarray}\label{EqBoltzmann}
\frac{dY}{dx}=- \frac{s \, \langle \sigma \, v_{r} \rangle}{x \, H(m_{\chi})} \, \left( \, Y^{\, 2}- Y_{EQ}^{\, 2} \, \right) \; ,
\end{eqnarray}
where $x:=m_{\chi}/T$ is the universe temperature normalized by the DM mass $m_{\chi}$, $H(m_{\chi})$ is the Hubble parameter when $T=m_{\chi}$, $Y$ is the
ratio of the DM number density divided by the entropy density $(s)$, $Y_{EQ}$ is the similar one when the particles are in thermal in equilibrium, and
$\langle \sigma \, v_{r} \rangle$ is the thermal average of the DM annihilation cross section $(\sigma)$ times the relative velocity $(v_{r})$. Furthermore, we list below some important formulas
\begin{eqnarray}\label{functions}
s = \frac{2\pi^2}{45} \, g_{\ast} \, \left( \frac{m_{\chi}}{x} \right)^{3}
\hspace{0.2cm} , \hspace{0.2cm}
H(m_{\chi}) = \sqrt{\frac{4\pi^3}{45} \, g_{\ast}} \, \frac{m_{\chi}^{\, 2}}{M_{Pl}}
\hspace{0.2cm} , \hspace{0.2cm}
s \, Y_{EQ} = \frac{g_{\chi}}{2\pi^{2}} \, \frac{m_{\chi}^{\, 3}}{x} \, K_{2}(x) \; ,
\end{eqnarray}
where $M_{Pl}=1.22 \times 10^{19} \, \mbox{GeV}$ is the Planck mass,
$g_{\chi}=1$ is the number of degrees of freedom for the scalar DM, $g_{\ast}=106.75$ (for SM particles)
is the effective total number of degree of freedom for the particles in thermal equilibrium, and $K_{2}$
is the modified Bessel function of the second kind.
The relic density is defined, with good approximation, by the asymptotic solution of the Boltzmann equation
\begin{eqnarray}
\Omega_{DM}h^2 =\frac{m_{\chi} \, s_{0} \, Y (\infty)}{\rho_{c}/h^2} \simeq
\frac{1.07\times 10^{9} \, X_{f}}{\sqrt{g_{\ast}} \, M_{Pl} \, \langle \sigma \, v_{r} \rangle} \; ,
\end{eqnarray}
where $s_{0}=2896 \, \mbox{cm}^{-3}$ is the entropy density in the present universe, $\rho_{c}/h^2=1.05 \times 10^{-5} \, \mbox{GeV}/\mbox{cm}^3$
is the critical density, and $Y(\infty)$ is the asymptotic solution of the Boltzmann equation. In this approximation, $X_{f}=\ln(X)-0.5 \, \ln(\ln(X))$, in which $X$ is given by
\begin{eqnarray}
X=0.038 \, \sqrt{ \, \frac{g_{\chi}}{g_{\ast}} \, } \, M_{Pl} \, m_{\chi} \, \langle \sigma \, v_{r} \rangle \; .
\end{eqnarray}
For the case (i), the thermal cross section times the relative velocity is \cite{HanWanEPJC2018}
\begin{eqnarray}\label{crosssecff}
\langle \sigma \, v_{r} \rangle \simeq \frac{2g_{X}^{4}}{27\pi\tan^4\theta_{W}}
\frac{ m_{\chi}^2 \, v^2}{(4m_{\chi}^2-m_{Z^{\prime}}^2)^2+m_{Z^{\prime}}^2 \, \Gamma_{Z'}^{2}}
\sum_{f} N_{f}^{c} \left( \, |c_{V}^{(f)}|^2 \, + \, |c_{A}^{(f)}|^{\, 2} \, \right) \; ,
\end{eqnarray}
where we have neglected the SM fermion mass in relation to DM mass $m_{\chi}$. The total $Z'$ decay width is
\begin{equation}
\Gamma_{Z'}=\frac{ m_{Z'} }{36\pi}
\frac{g_{X}^{\, 2}}{\tan^{2}\theta_{W}} \sum_{f} N_{c}^{(f)} \left( \, |c_{V}^{(f)}|^2 + |c_{A}^{(f)}|^2 \, \right)
+\frac{m_{Z^{\prime}}}{48\pi}\frac{g_{X}^2}{\tan^{2}\theta_{W}}\left(1-\frac{4m_{\chi}^2}{m_{Z'}^2} \right)^{3/2} \! \Theta(m_{Z^{\prime}}-2m_{\chi}) \; ,
\end{equation}
in which the $Z^{\prime}$ decay mode into DM pair $\chi^{\star} \, \chi$ is active for the condition $m_{Z^{\prime}} > 2 \, m_{\chi}$.
The relic density is plotted as function of the DM scalar mass $m_{\chi}$ in the figure (\ref{relic331}) for $m_{Z^{\prime}}=5.36$ TeV
and the $U(1)_{X}$ coupling constant $g_{X}=0.36$ constraints by the parametrization in (\ref{chargee}).
The DM mass is assumed at the range $0.5$ TeV to $5.5$ TeV. The dashed line sets the actual DM observed relic density at $\Omega_{DM}h^2=0.12 \, \pm \, 0.0012$ \cite{Planck2018}. The intersection of DM abundance with the black curve yields the two solutions for DM masses : $m_{\chi}=1.701$ TeV and $m_{\chi}=4.248$ TeV. If $Z^{\prime}$ decays into DM pair $\chi^{\ast} \, \chi$, the first solution at $m_{\chi}=1.701$ TeV is kinetically allowed for the $Z^{\prime}$ mass of $5.36$ TeV. The minimum peak in the relic density curve is at $m_{\chi}=2.70$ TeV, that is approximately $m_{Z^{\prime}} \approx 2 \, m_{\chi}$.
For the second case (ii), the thermal cross section of $\chi^{\ast} \, \chi \rightarrow Z^{\prime} Z^{\prime}$ has the result \cite{HanWanEPJC2018} :
\begin{eqnarray}\label{crosssecZZ}
\langle \sigma \, v_{r} \rangle=\frac{g_{X}^{4}}{11664\pi \tan^4\theta_{W}} \, \frac{m_{\chi}^2}{m_{Z^{\prime}}^4}\left(1-\frac{m_{Z^{\prime}}^2}{m_{\chi}^2} \right)^{1/2} \!\! \left(1-\frac{m_{Z^{\prime}}^2}{2m_{\chi}^2} \right)^{\!-2} \times
\nonumber \\
\times \left[4\left(1-\frac{m_{Z'}^2}{m_{\chi}^2} \right)^{2}-\left(4-4 \, \frac{m_{Z^{\prime}}^2}{m_{\chi}^2}-3 \, \frac{m_{Z^{\prime}}^4}{m_{\chi}^4}\right)\left(1-\frac{m_{Z'}^2}{2m_{\chi}^2} \right)^{2} \right] \; ,
\end{eqnarray}
that satisfies the condition $m_{\chi}>m_{Z^{\prime}}$. When $m_{\chi}>m_{Z^{\prime}}$, the cross section (\ref{crosssecZZ})
independent velocity dominates over the cross section dependent velocity from (\ref{crosssecff}).
For $m_{\chi} \gg m_{Z^{\prime}}$, the cross section (\ref{crosssecZZ}) is proportional to the inverse of squared DM mass :
\begin{eqnarray}
\left. \langle \sigma \, v_{r} \rangle \right|_{m_{\chi} \gg m_{Z^{\prime}}} \simeq \frac{g_{X}^{4}}{5832\pi \tan^4\theta_{W}} \frac{1}{m_{\chi}^2}
\simeq \frac{10^{-5}}{m_{\chi}^2} \; .
\end{eqnarray}
In this case, the abundance relic density $0.12$ is reproduced when $m_{\chi}=31.62$ GeV, and consequently, we have the scenario of a light $Z^{\prime}$ DM portal.
The direct detection for the DM mass at the $\sim 1$ TeV order is bounded by the PANDAX2017. It has the upper bound for the
spin independent (SI) cross section of $\sigma^{SI} \lesssim 10^{-9}$ pb \cite{PANDAX2017}. The XENON1T(2t.y) has the limit at
one order below $\sigma^{SI} \lesssim 10^{-10}$ pb \cite{Xenon20172ty}. The SI cross section
for nucleon-scalar DM process $\chi \, N \rightarrow \chi \, N$ is given by the expression :
%
%
\begin{eqnarray}
\sigma^{SI}\simeq \frac{1}{81\pi} \frac{g_{X}^4}{\tan^{4}\theta_{W}} \frac{\mu_{N}^{2}}{m_{Z^{\prime}}^{\, 4}}
\simeq 2.413 \times 10^{-7} \,
\left( \frac{\mu_{N}}{1 \, \mbox{GeV}} \right)^{2} \left( \frac{1 \, \mbox{TeV}}{m_{Z^{\prime}}} \right)^{4} \, (\mbox{pb}) \; ,
\end{eqnarray}
where $\mu_{N}=m_{N} \, m_{\chi}/(m_{N} + m_{\chi})$ is the reduced mass of the system nucleon-DM, and $m_{N}=939$ MeV is the averaged nucleon mass.
When $m_{Z^{\prime}}=6.89$ TeV and $m_{\chi}=1$ TeV, the SI cross section has the same order from XENON1T(2t.y) bound :
$\sigma^{SI} \lesssim 10^{-10}$ pb. The LUX experiment for DM detection is expected with a bound of one lower
in relation to the XENON1T : $\sigma^{SI} \lesssim 3 \times 10^{-11}$ pb for a DM mass of $1$ TeV.
\begin{figure}[h]
\centering
\includegraphics[width=0.77\textwidth]{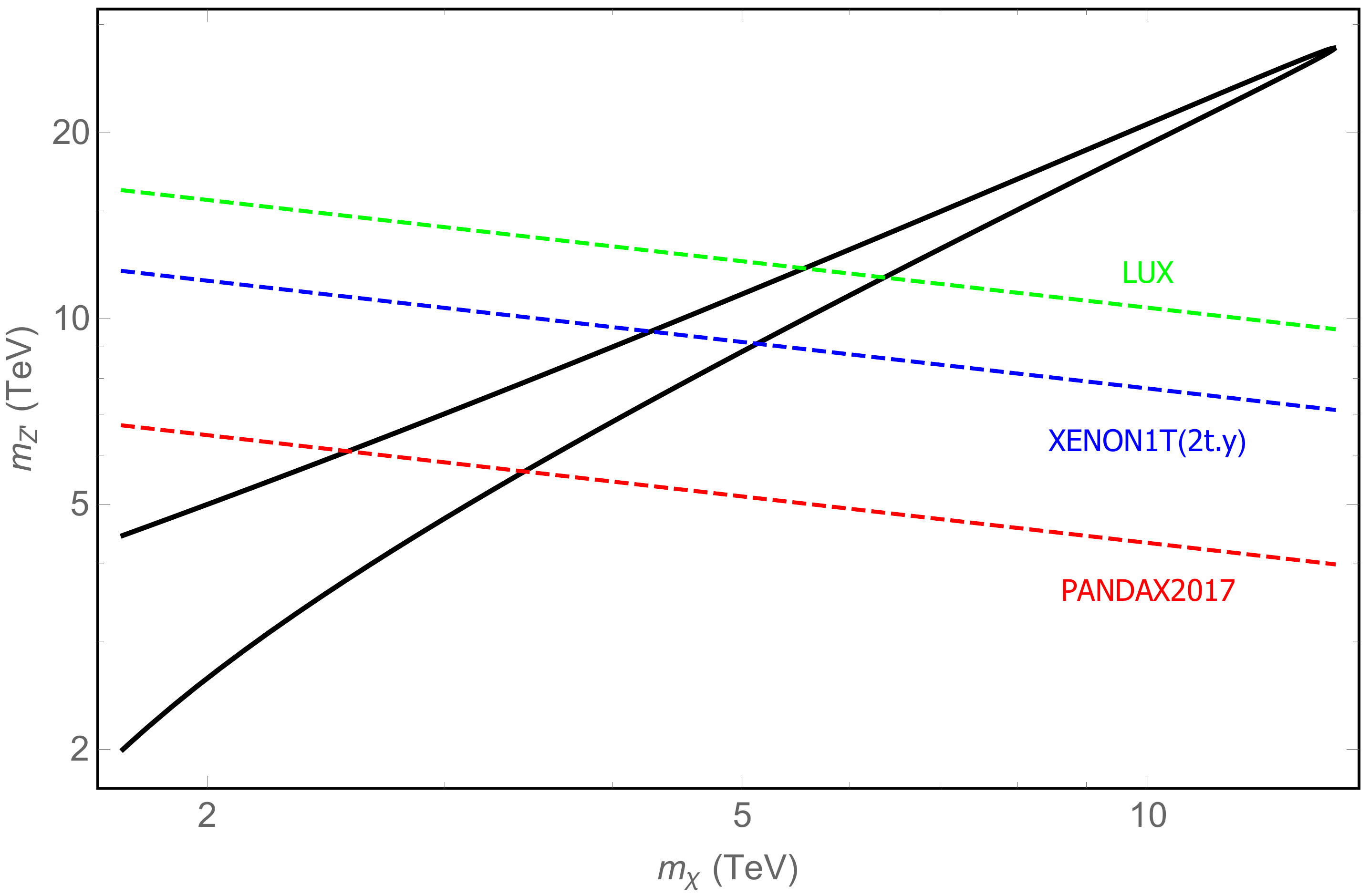}
\caption{
The $Z^{\prime}$ mass as function of the DM mass. The abundance relic density is reproduced along
the black line. The dashed lines are the bounds of PANDAX2017 (red), XENON1T(2t.y) (blue) and LUX (green). }
\label{bounds}
\end{figure}
%

%
%

%
The plot in (\ref{bounds}) shows the parameter space of the DM mass $m_{\chi}$ versus the $Z^{\prime}$ mass,
when the $U(1)_{X}$ coupling constant is $g_{X}=0.36$. In this analysis, we also consider the $Z^{\prime}$ mass
as a free parameter. Along the black line the abundance relic density $\Omega_{DM} . h^2=0.12$ is reproduced.
The dashed lines are the bounds of PANDAX2017 (dashed red line), XENON1T(2t.y) (dashed blue line) and the LUX experiment
(Green dashed line), respectively.
%

%
\section{conclusions}
\label{sec7}
The gauged model $SU(3)_{c} \times SU_{L}(3) \times U(1)_{X}$ with a discrete $\mathbb{Z}_{2}$ symmetry is studied in this paper
to investigate the extra CP violation in the mass basis of the fermions. The discrete symmetry makes a stable scalar field in the
Higgs sector, that is a possible candidate to the DM content. The $Z^{\prime}$ gauge boson in 331 model is the DM portal to link
the SM fermions with the scalar DM candidate.
The relic density is calculated for the process DM anti-DM annihilated into SM fermion anti-fermion pair through the $Z^{\prime}$ portal,
in the fig.(\ref{relic331}). The parameter space of DM mass versus $Z^{\prime}$ mass shows the allowed region to constraint the
observed relic abundance $0.12$ along the black line in the fig.(\ref{bounds}). It is also included the bounds of PANDAX2017, XENON1T(2t.y),
and LUX experiment on this parameter space.
The model also includes a new sector of fermions with masses bounded at the TeV scale that have properties similar to the quarks in the SM.
In the mass basis, these exotic fermions couples with the quarks of the SM mediated by the charged gauge boson $Y$ of the 331
model bounded with mass of $4.4$ TeV through the LHC constraints. The new couplings introduce mixing angles and extra phases that violate CP
and can lead to a study of a electroweak bariogenesis in the 331 model. Furthermore, a new phenomenology for the decay modes of $Y$
in which indicates an existence of asymmetry between matter and anti-matter that can be investigate in a future research paper.

\section*{Acknowledgments}
The work of M. J. Neves has been supported by the Conselho Nacional de Desenvolvimento Cient\'ifico e Tecnol\'ogico (CNPq) under grant
313467/2018-8 (GM). M. J. Neves would like to thanks the Department of Physics \& Astronomy at the University of Alabama for the hospitality
during his visit as a J-1 Research Scholar.
%

%


%
%

%



%



%
\end{document}